\documentclass[times,11pt,a4paper]{article} 

\makeatletter

\def\@thmcountersep{}
\def\@thmcounterend{:}

\def\mynewtheorem{\@ifstar{\@sthm}{\@Sthm}}

\def\@spnthm#1#2{%
  \@ifnextchar[{\@spxnthm{#1}{#2}}{\@spynthm{#1}{#2}}}
\def\@Sthm#1{\@ifnextchar[{\@spothm{#1}}{\@spnthm{#1}}}

\def\@spxnthm#1#2[#3]#4#5{\expandafter\@ifdefinable\csname #1\endcsname
   {\@definecounter{#1}\@addtoreset{#1}{#3}%
   \expandafter\xdef\csname the#1\endcsname{\expandafter\noexpand
     \csname the#3\endcsname \noexpand\@thmcountersep \@thmcounter{#1}}
   \expandafter\xdef\csname #1name\endcsname{#2}%
   \global\@namedef{#1}{\@spthm{#1}{\csname #1name\endcsname}{#4}{#5}}
                              \global\@namedef{end#1}{\@endtheorem}}}

\def\@spynthm#1#2#3#4{\expandafter\@ifdefinable\csname #1\endcsname
   {\@definecounter{#1}%
   \expandafter\xdef\csname the#1\endcsname{\@thmcounter{#1}}
   \expandafter\xdef\csname #1name\endcsname{#2}
   \global\@namedef{#1}{\@spthm{#1}{\csname #1name\endcsname}{#3}{#4}}
                               \global\@namedef{end#1}{\@endtheorem}}}

\def\@spothm#1[#2]#3#4#5{%
  \@ifundefined{c@#2}{\@latexerr{No theorem environment `#2' defined}\@eha}
  {\expandafter\@ifdefinable\csname #1\endcsname
  {\global\@namedef{the#1}{\@nameuse{the#2}}
  \expandafter\xdef\csname #1name\endcsname{#3}
  \global\@namedef{#1}{\@spthm{#2}{\csname #1name\endcsname}{#4}{#5}}
  \global\@namedef{end#1}{\@endtheorem}}}}

\def\@spthm#1#2#3#4{\topsep 7\p@ \@plus2\p@ \@minus4\p@
\refstepcounter{#1}
\@ifnextchar[{\@spythm{#1}{#2}{#3}{#4}}{\@spxthm{#1}{#2}{#3}{#4}}}

\def\@spxthm#1#2#3#4{\@spbegintheorem{#2}{\csname the#1\endcsname}{#3}{#4}
                    \ignorespaces}

\def\@spythm#1#2#3#4[#5]{\@spopargbegintheorem{#2}{\csname
       the#1\endcsname}{#5}{#3}{#4}\ignorespaces}

\def\@spbegintheorem#1#2#3#4{\trivlist
                 \item[\hskip\labelsep{#3#1\ #2\@thmcounterend}]#4}

\def\@spopargbegintheorem#1#2#3#4#5{\trivlist
      \item[\hskip\labelsep{#4#1\ #2}]{#4(#3)\@thmcounterend\ }#5}

\def\@sthm#1#2{\@Ynthm{#1}{#2}}

\def\@Ynthm#1#2#3#4{\expandafter\@ifdefinable\csname #1\endcsname
   {\global\@namedef{#1}{\@Thm{\csname #1name\endcsname}{#3}{#4}}
    \expandafter\xdef\csname #1name\endcsname{#2}%
    \global\@namedef{end#1}{\@endtheorem}}}

\def\@Thm#1#2#3{\topsep 7\p@ \@plus2\p@ \@minus4\p@
\@ifnextchar[{\@Ythm{#1}{#2}{#3}}{\@Xthm{#1}{#2}{#3}}}

\def\@Xthm#1#2#3{\@Begintheorem{#1}{#2}{#3}\ignorespaces}

\def\@Ythm#1#2#3[#4][#5]{\@Opargbegintheorem{#1}
       {#4}{#2}{#3}{#5}\ignorespaces}

\def\@Begintheorem#1#2#3{#3\trivlist
                           \item[\hskip\labelsep{#2#1\@thmcounterend}]}

\def\@Opargbegintheorem#1#2#3#4#5{#4\trivlist
      \item[\hskip\labelsep{\indent #3#2}]{#3#1 (#5)\@thmcounterend\ }}

\makeatother

\usepackage[cmex10]{amsmath}
\usepackage{amssymb}
\usepackage{amsthm}
\usepackage{indentfirst} 
\usepackage{amsmath} 
\usepackage{latexsym} 
\usepackage{graphics} 
\usepackage{graphicx} 
\usepackage{mathrsfs}
\usepackage{nextpage}
\usepackage{times} 
\usepackage{amsfonts}
\usepackage{hhline}
\usepackage{url,cite}
\usepackage{tabularx}
\usepackage{mathrsfs}
\usepackage{array}
\usepackage{algorithmic}
\usepackage{mdwmath}
\usepackage{mdwtab}
\usepackage{eqparbox}
\usepackage{xcolor}
\usepackage{epsfig, graphics,subfigure,delarray}

\newtheorem{theorem}{Theorem}
\newtheorem{definition}{Definition}
\newtheorem{lemma}[theorem]{Lemma}

\newtheorem{result}[theorem]{Result}

\newcommand\mycleardoublepage{\cleartooddpage[\thispagestyle{empty}]}

\def\keywords{\normalfont
    \if@twocolumn
      \small\bfseries\textit{Index Terms}---\,\relax
    \else
      \begin{center}\small\bfseries Index Terms\end{center}\quotation\small
    \fi}
\def\endkeywords{\relax\vspace{0.67ex}
    \par\if@twocolumn\else\endquotation\fi
    \normalsize\normalfont}

\begin{document} 
\pagenumbering{roman}

%\thispagestyle{empty}
%\mycleardoublepage
\begin{center}\fontsize{14}{12}{\textbf{Resource Allocation for Improved User Experience with Live Video Streaming in 5G}} \end{center}

%\vspace{+2cm}

\begin{center} 
 Fidan Mehmeti and Thomas F. La Porta
\end{center}

%\vspace{+2cm}

\begin{abstract}
Providing a high-quality real-time video streaming experience to mobile users is one of the biggest challenges in cellular networks. This is due to the need of these services for high rates with low variability, which is not easy to accomplish given the competition among (usually a high number of) users for constrained network resources and the high variability of their channel characteristics. %Buffers can only partially mitigate the deterioration caused by the variability in the network throughput. %that is not enough.  
%This is due to the need for high constant rates at all times for this type of services, which is not easy to achieve given the limited network resources and the competition among (usually a high number of) mobile users. 
 A %better 
 way of improving the user experience is by exploiting their buffers and the ability to provide a constant data rate to everyone, as one of the features of 5G networks. \textcolor{black}{However, the latter is not very efficient.} %and by reallocating the unused resources to the same users. 
 To this end, in this paper we provide a theoretical-analysis framework for resource allocation in 5G networks that leads to an optimized performance for %real-time 
\textcolor{black}{live} video streaming. We do this by solving three problems, in which the objectives are to provide the highest %possible 
\textcolor{black}{achievable}
video resolution to \textcolor{black}{all} one-class and two-class users, and to 
 maximize the number of users that experience 
 \textcolor{black}{a given} %possible 
 resolution. The %theoretical 
 analysis is validated by simulations that are run on trace data. \textcolor{black}{We also compare the performance of our approach against other techniques for different QoE metrics.}
 Results show that \textcolor{black}{the} performance can be improved %significantly with our approach. 
 by at least $15$\% with our approach.  
\end{abstract}


\thispagestyle{empty}
\mycleardoublepage

\tableofcontents
\pagebreak
%\listoftables
\listoffigures
\pagebreak

\pagenumbering{arabic}

\section{Introduction}
\label{sec:intro}
%The explosive growth in the number of applications, such as video streaming~\cite{5G-vision}, augmented reality~\cite{Elbamby18}, etc., and their stringent service requirements~\cite{nokia5g}, 
%has made the LTE networks inept to provide a high Quality of Experience (QoE) to mobile users interested in these applications/services. 

LTE networks are unable to provide a high Quality of Experience (QoE) to mobile users interested in increasingly popular applications/services, such as video streaming~\cite{5G-vision}, augmented reality~\cite{Elbamby18}, etc., because of their stringent service requirements~\cite{nokia5g}.   
A better solution to overcome this problem than increasing the density of current \textcolor{black}{(4G)} network infrastructure~\cite{qi5G}, which comes at a considerable cost \textcolor{black}{with modest gains}, is the deployment of the next generation of cellular networks, 5G. The possibility of network slicing in 5G\cite{p11} enables assigning \emph{dedicated} network resources to the same type of service, e.g., users watching live the same event.

One of the services that poses serious strains on cellular network operation is video streaming \textcolor{black}{of live events}. Providing a high-quality real-time video streaming experience %\textcolor{black}{of a live event}   
to mobile users, expressed through a consistently high video quality with seldom rebuffering events \textcolor{black}{and no (or a small number of) packets dropped}, is not  straightforward. The reason for that is the need for \emph{high data rates with low variability}~\cite{nokia5g}. This requirement is particularly challenging to accomplish because of the competition for limited network resources among the ever increasing number of users running bandwidth-hungry applications on their smartphones, and their channel characteristics that exhibit high variability.    

According to~\cite{Nam16}, the \emph{variable playout rates} %\textcolor{red}{are the second cause of abandoning video streaming, with up to $21$\% of users abandoning the video, 
%with}  
\textcolor{black}{cause up to $21$\% of users abandoning video streaming, which is the second leading cause;} 
the main cause is the \emph{rebuffering (outage) events}. Furthermore, Mux (mux.com) conducted a survey with more than 1000 US respondents~\cite{mux} about their experience\textcolor{black}{s} with video streaming services. According to that survey, the \emph{low picture quality (low video resolution) was the most frustrating problem for $14.3$\% of the viewers}, and $57$\% of respondents  had abandoned a video in the past due to low video resolution. Hence, %the importance of not only minimizing 
it is important to not only minimize 
the video outage due to rebuffering events, but also to provide consistently \textcolor{black}{a} high video resolution.

%The explosive growth in the number of applications and services, such as video streaming~\cite{5G-vision}, augmented reality~\cite{Elbamby18}, etc., and their stringent service requirements 

A way of improving the user experience is by jointly exploiting their buffers on end devices (that amortize the data rate variability), the ability to provide a \textcolor{black}{constant data rate} to everyone in 5G~\cite{ericsson-5g}, and the proper allocation of network resources. While providing a constant data rate with a low outage in 5G, known as \emph{consistent rate}, can indeed mitigate the variations in video resolution, it has been shown~\cite{Fidan_TMC} that it is quite inefficient in terms of resource utilization and has been described as an ``expensive feature''. %\textcolor{red}{In this paper, we show that because of the nature of the traffic in \textcolor{black}{live} video streaming, having a sufficiently large buffer mitigates the need for strict consistent data rates, and even optimizes the performance with proper resource allocation \textcolor{black}{by allowing a slightly higher delay, which does not exceed $20$\:s with some policies.}}  
\textcolor{black}{In this paper, we show that because of the nature of the traffic in live video streaming, having even a small buffer (so that a given traffic latency is not exceeded) mitigates the need for strict consistent data rates, and even optimizes the performance with proper resource allocation, by allowing only a small number of packets to be dropped.}
%\textcolor{black}{Our results show that we can achieve constant playout rates with a stream latency not exceeding $20$\:s  with some policies.}  
Hence, our focus in this work is to efficiently allocate the resources so that the user's QoE is \textcolor{black}{optimized}. %improved.  

There are several questions that arise when dimensioning network slices or when designing admission control policies for real-time video applications in 5G. First and foremost, what is the maximum resolution at which a user can watch a live event on her smartphone given her channel conditions and the number of other users in the cell? It is also of interest to know the amount of resources a user needs in order to play the video at the lowest \textcolor{black}{acceptable} resolution. %What about the resources needed to experience the best video quality? 
Next, given the competition for the network resources, what allocation scheme enables all the users to experience the same video quality, and what resolution is that? Another important question is how to allocate resources so that the number of users with ultimate video experience \textcolor{black}{(or another video resolution)} 
is maximized. %Finally, deciding on how to allocate network resources to provide proportional fairness in playout rates is another interesting problem.    

%The following research questions arise when dimensioning network slices for real-time video applications in 5G networks:
%\begin{itemize}
%\item What is the maximum resolution that a user can watch a live event in her smart phone given her wireless conditions and the number of other users in the cell?
%\item If a small outage in providing a playout rate is allowed, by how much the video resolution can be improved?
%\item How to perform the network resource allocation so that a maximum number of users could playout their videos at the highset possible resolution, whereas the other users playout their videos at a minimum acceptable resolution?
%\item If the goal is to provide the maximum possible playout rate with a given outage, how should the base station allocate its resources to all the users in the cell?
%\item How should the resources be assigned if the objective is to provide proportional fairness in playout rate?
%\end{itemize}
% The possibility of network slicing in 5G\cite{p11} enables assigning \emph{dedicated} network resources to the same type of services, i.e., consistent users in our scenario.
%\emph{Network slicing, transmission model with BTS from the MobiHoc paper}
In this paper, to address the aforementioned issues, we formulate %and solve 
three %optimization 
problems 
whose solutions lead to efficient resource allocation %approaches 
\textcolor{black}{policies} to be followed by the cellular operator, resulting in significant QoE improvements for mobile users. Our study is particularly important because it can help cellular operators efficiently allocate network resources depending on the optimization objective, \textcolor{black}{while providing a constant playout rate throughout the entire streaming.} 
%\textcolor{black}{Also, it shows that the frequency of rebuffering events can be decreased significantly without deteriorating the video resolution considerably},
\textcolor{black}{Also, it shows that the loss of information is very low even with very small buffers, thus maintaining a small delay between the occurrence of the event and the time it is played out on the smartphone.}  % preserving \emph{freshness of information}},
The approach is flexible enough to be implemented on top of any of the proposed 5G architectures~\cite{trivisonno2015towards},~\cite{gupta2015survey}.   

Specifically, our main contributions are:
\begin{itemize}
%\item \textcolor{blue}{Mention: We provide a model that by knowing only the distribution of the per-block rates of the users, can predict what playout rate can be provided with a maximum outage and drop rate.}
\item We model the real-time video streaming on smartphones as a discrete-time queueing process \textcolor{black}{with finite buffer and fixed playout rate} and derive the maximum constant playout rate \textcolor{black}{(resolution)} that a user can experience for $1-\epsilon$ of the time \textcolor{black}{with a maximum number of packets than can be dropped}, given its channel conditions and network resources available.
\item We derive the resource allocation that maximizes the video resolution that can be played out by all the users with the same outage \textcolor{black}{and drop rate}.
\item We consider the problem of maximizing the number of users that can be provisioned with \textcolor{black}{a target} video experience while providing the minimum acceptable video resolution to all the other users, and derive the resource allocation that provides that.
%\item We also consider how to allocate the network resources in order to provide \emph{proportional fairness} in playout rates (video resolution) of all the users of interest.
%\item We validate our analysis \textcolor{black}{via} extensive synthetic simulations and obtain several engineering insights. In particular,
%we show that taking a small risk (in terms of not delivering the committed data rate) provides higher consistent rates.
\item We validate \textcolor{black}{the} analysis through simulations run on trace data and also obtain several engineering insights, \textcolor{black}{such as the need for very small user's buffers even for highly variable channel conditions.} %to validate the approximation,  
%and show that even in those cases our theory is able to predict the performance quite accurately. 
\item \textcolor{black}{We show that our approach of constant playout rate provides a better user experience than the adaptive bit rate (ABR) techniques for different QoE metrics.}
\end{itemize}

The remainder of this paper is organized as follows. The %system 
model and problem formulation are presented in Section~\ref{sec:model}. This is followed by the theoretical analysis in Section~\ref{sec:analysis}. We solve three problems in Section~\ref{sec:optimization}. \textcolor{black}{We present a benchmark model and the QoE metrics in Section~\ref{sec:qoe}.} 
The performance evaluation together with some engineering insights are presented in Section~\ref{sec:evaluation}. We discuss some related work in Section~\ref{sec:related}. Finally, Section~\ref{sec:conclusion} concludes the work.
  
\section{Performance modeling}
\label{sec:model}

\subsection{System model}
\begin{figure*}[t]
\centering
\includegraphics[width=\textwidth]{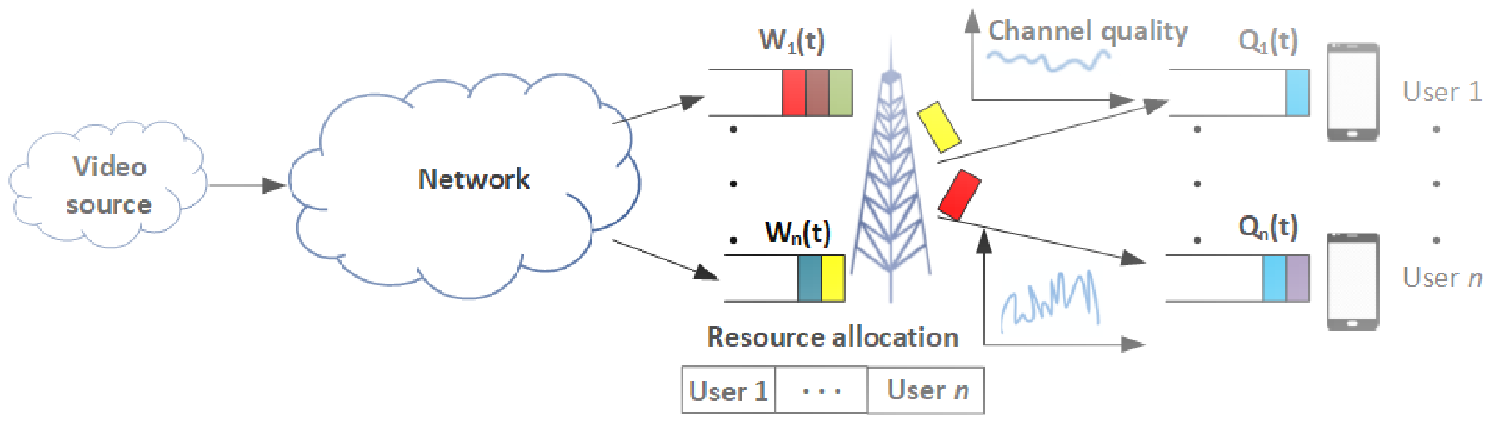}
%\vspace{-9pt}
\caption{\textcolor{black}{Illustration of the real-time video streaming process in a cell with \emph{n} users}.}
\label{system_model}
%\vspace{-10pt}
\end{figure*}

We consider mobile users within the coverage area of a \textcolor{black}{5G} macro base station (\textcolor{black}{gNodeB}) \textcolor{black}{in the} %licensed 
sub-6 GHz band %\textcolor{black}{frequencies} %cellular network, 
with the focus on the downlink (Fig.~\ref{system_model}). We assume that the users watch the same event live on their smartphones, e.g., a football match, or \textcolor{black}{a show.}  

As in LTE networks, the block resource allocation scheme is used in 5G as well, with \emph{physical resource blocks (PRB)} being the unit of allocation~\cite{block}. The frame duration is $\Delta t$. Within a frame, different blocks are assigned to different users. In general, the assignment will change in every frame.  So, there is a scheduling \emph{both in time and frequency}. 
The total nuFmber of blocks \textcolor{black}{dedicated to this use case} is $K$. 

Users will experience different channel characteristics in different blocks (different frequencies) even within the same frame, and hence a different per-block \textcolor{black}{Signal-to-Interference-plus-Noise Ratio} (SINR). \textcolor{black}{The latter is a function of the base station transmission power of the cell where the users are, the transmission power of neighboring cells transmitting on the same frequencies (inter-cell interference), AWGN noise and the corresponding channel gains~\cite{Fidan_TMC}.} Due to user's mobility and time-varying channel characteristics, per-block SINR changes from one frame to another even for the same block. This varying \emph{per-block} SINR translates into a varying \emph{per-block rate}, which is obtained using a modulation and coding scheme (MCS). In our system, we consider an %adaptive 
MCS with $m$ possible values (the typical value of $m$ is either $15$ or $31$)~\cite{3GPP_5G_NR}. %,~\cite{AMC2}.  
%The mobile users will see a time-varying per-channel SINR due to their mobility and time-varying channel characteristics, and as a consequence of that, a varying \emph{per-channel rate}. 
%This data rate is a function of the per-channel SINR and of the modulation and coding scheme, where the latter is used to translate the SINR into a per-channel rate. In our case, we consider the adaptive modulation and coding scheme (MCS) with $m$ possible values (the typical value of $m$ is 15)~\cite{AMC,AMC2}. 
E.g., if the per-block SINR lies in the interval $[\gamma_j,\gamma_{j+1}]$, with $\gamma_j$ and $\gamma_{j+1}$ being the thresholds of the %adaptive 
MCS ($j=1,\ldots,m$), the per-block rate in that frame would be $r_j$~\cite{amc_2}. For every user, we assume flat channels \textcolor{black}{(blocks)} in a frame, i.e., the per-block rate does not change during the frame, but it changes from one frame to another randomly\footnote{\textcolor{black}{In Section~\ref{sec:evaluation}, we show that even when per-block rates of a user in contiguous frames are correlated, the analysis provides a close match \textcolor{black}{to} %between the theory and 
actual results.}}.  

To maintain tractability, 
we make a simplifying assumption. %\textcolor{black}{in order to keep the analysis tractable}. 
%Namely, 
\textcolor{black}{We assume that the base station transmission power and channel characteristics of a user are the same across all $K$ blocks in a frame (flat fading)}. This way our problem reduces to a one-dimensional scheduling (in time). So, instead of deciding how many and which blocks to assign to every user, we use another parameter, which is defined as:
\textcolor{black}{
\begin{definition}
The ratio of frame for which all the blocks are allocated to user $i$ is called \textbf{frame ratio}. It is denoted by $Y_i$, and can take values in the interval $[0,1]$. 
\end{definition}}

The blocks are assigned orthogonally during the frame, so that no two users receive them simultaneously.  
%we decide on the corresponding ratio of the frame (\emph{frame ratio}), $Y$, all the  blocks are going to be assigned \textcolor{black}{(orthogonally in time)} to everyone. 
\textcolor{black}{This is a reasonable simplification as frame ratio can be translated into the corresponding number of PRBs.} %Since our focus here is on \emph{reallocation}, this assumption is reasonable as this \emph{frame ratio} could be translated into the corresponding resource blocks.  The second reason to do so is because that is the \textcolor{black}{only} way to have a tractable theoretical analysis. Hence, 
Hence, from now on, we will be considering only the frame ratio, \textcolor{black}{with values in $[0,1]$}, %as the unit of resource allocation.
as the resource allocation unit.

%the transmission power is allocated equally among all the $K$ channels (blocks) and that the channel gain for a user is the same over all channels in a frame.

As a result of previous assumptions, in every frame user's $i$ per-block rate can be modeled as a discrete random variable, $R_{i}$, with values in $\left\{r_{1},r_{2},\ldots,r_{m}\right\}$, such that $r_{1}<r_{2}<\ldots<r_{m}$, %with a cumulative distribution function $F_{R_{i}}(x)$.
\textcolor{black}{with a probability mass function (PMF) $p_{R_{i}}(x)$, which is a function of  user's $i$ SINR over time.\footnote{\textcolor{black}{We assume that based on the history of changes in SINR for a user, the base station knows the per-block rate probabilities for every user.}}}  
%\footnote{The per-channel data rate is clearly a function of time, but for simplicity we will omit the time dependency notation-wise only.} 

\emph{Number of users:} The number of (mobile) users in the cell is $n$.
%We consider the case of heterogeneous users, i.e., users that have different 
The users have different 
per-block rate distributions. %First, we assume that users are always active in the cell, and then consider the scenario in which users are not always active. 

\emph{Video content:} The content of the video, after being generated, ``travels'' through the core of the network to the corresponding base station (Fig.~\ref{system_model}). The video content of every user is associated with two buffers. \textcolor{black}{The first is on} the base station side where the video packets\footnote{Note that our model is general and instead of packets, the granularity \textcolor{black}{level} can be increased to chunks, without affecting the theoretical analysis.} %Hence, throughout the paper, we will be considering packets to be the unit size of information.} 
arriving from the core of the network are temporarily stored. These video packets are then transmitted to the user, and are to be stored in the second buffer, from where \textcolor{black}{they} %the packets 
will be played out. \textcolor{black}{The time interval between the packet being generated and its playout is called \emph{streaming latency}.}  
It is well known that the backhaul bandwidth is much larger than the bandwidth between the base station and mobile users. Hence, the latter is the bottleneck of the network in terms of delay, and that will be \textcolor{black}{the part of network of our interest.} %with the latency referring to it}. 

\emph{Playout rate:} The rate at which the video content is played on the smartphone of a user is called \emph{playout rate} or \emph{bitrate}, and is denoted by $U$. It should not be confused with the data rate from the network. The former ``plays'' the packet, whereas the latter ``brings'' \textcolor{black}{it} to the user. \textcolor{black}{This is one of our QoE metrics of interest.} \textcolor{black}{It is directly related to the \emph{video resolution}. The higher the playout rate, the higher the video resolution is. E.g., playing videos in the resolution range 144p-1080p requires a playout rate of $0.1-5.8$\:Mbps~\cite{guohong}. The ultimate 4K UHD videos require up to $40$\:Mbps~\cite{qi5G}.}  

%\emph{Video resolution:} There are $l$ levels of the video resolution $f_1,\ldots,f_l$, %where $f_1$ is the resolution with the lowest quality (\emph{minimum resolution} from now on), whereas $f_l$ is the resolution of the highest possible quality (\emph{maximum resolution} from now on). 
%\textcolor{black}{in increasing order of quality}. 
%The video resolution is proportional to the playout rate, i.e., the higher the playout rate, the higher the video resolution is. E.g. playing videos in the resolution range 144p-1080p requires a playout rate of 0.1-5.8 Mbps~\cite{guohong}. The ultimate 4K UHD videos require 40 Mbps~\cite{qi5G}.  %Hence, the paramount importance of providing as high a value of $K$ as possible. \emph{Cite Guohong's paper}

Since, as mentioned in Section~\ref{sec:intro}, most users are not happy when the resolution of the video changes over time, we strive to provide \emph{a \textcolor{black}{constant} playout rate to every user at all times.} \textcolor{black}{Therefore, in this work, we assume that every user will have a constant playout rate.}  

\emph{Outage:} When there are no packets in user's buffer, the video will stall until new packets arrive. This is known as a \emph{rebuffering event} or as an \emph{outage}, which is one of the most important measures of QoE, \textcolor{black}{and is our metric of interest too}. The probability of having an outage in a frame is $\epsilon$. Obviously, this should be as low as possible. Note that there is a tradeoff involved between the playout rate and the outage. \textcolor{black}{In general, allowing a higher outage may lead to higher playout rates. This also depends on user's buffer size.} %, as will be explained in more detail in  Section~\ref{sec:analysis}}.  
%\textcolor{black}{This is our second QoE metric of interest.}  %\textcolor{red}{depends on user buffer size}   

\textcolor{black}{\emph{Dropped packets:} %The buffers are of finite size at all the users. 
We assume that all the users have the same (finite) buffer size. This assumption gives rise to another 
phenomena, that of \emph{information loss}. Namely, when the arriving packets to the user's buffer find it full, they will be dropped. That part of information will be lost for the user. To capture this effect quantitatively, we use the parameter \emph{packet drop rate}, $\delta$, which denotes the ratio of the lost packets\footnote{We will refer to this quantity simply as \emph{drop rate}.}. It can also be interpreted as the probability for a packet to be dropped. The goal is to provide as low a packet drop rate as possible. %Note that there is a tradeoff involved between the playout rate and drop rate. 
Obviously, the higher the playout rate, the lower the drop rate is and vice versa.} 

\textcolor{black}{The outage $\epsilon$ and drop rate $\delta$ are the main driving forces (together with the network resources and the channel conditions of the users) that determine the value of the playout rate. Namely, increasing the playout rate leads to a decrease in the drop rate but also increases the outage.}

%In order to increase the consistent rate, an operator can relax the constraint on delivering the QoS 100\% of the time. Instead, the mobile operator can commit to providing the consistent rate for $1-\epsilon$ of the time to all the consistent users, where $\epsilon$ is the \emph{outage probability}. The value of the outage probability must be very small (few percents).

%Finally, \textcolor{black}{it is worth mentioning that although when using the mmWave technology~\cite{mmwave} it is indeed easier to provide consistent rates, the area a base station covers is very small due to the propagation characteristics of 60 GHz waves. Hence, the number of users that can receive service in a cell is usually very small, too. Therefore, our focus in this paper is on licensed cellular bands.}

\subsection{Queueing model}
%The frame duration is $\Delta t$. 
The packet size is $\sigma$.\footnote{\textcolor{black}{As shown in~\cite{Sengupta17}, the packet sizes in video streaming applications exhibit very low variability. Nevertheless, in Section~\ref{sec:evaluation} we relax this condition and show that our analysis can predict quite accurately the performance even for variable packet sizes.}} \textcolor{black}{The maximum number of packets in the buffer is $B$.}  
If in frame $t$ the data rate of user $i$ is $C_i(t)$, the total number of packets that arrive \textcolor{black}{and are queued} \textcolor{black}{(if there is enough space)} in the buffer of user $i$ during that frame is $A_i(t)=\lfloor{\frac{C_i(t)\Delta t}{\sigma}\rfloor}$. \textcolor{black}{They are played out according to the First Come First Served (FCFS) order of service. The earliest that the packets arriving during frame $t$  can be played out is in frame $t+1$.}  
The rate $C_i(t)$ is a function of channel conditions (per-block rate) of user $i$ and of the total number of users in the cell, and is i.i.d. across frames, \textcolor{black}{implying that $A_i(t)$ is i.i.d. too.} \textcolor{black}{In case some of the arriving packets find the buffer full, they will be dropped.} 
%\footnote{For notational convenience we will omit the reference to $t$ from now on.} 
%Also, we assume that the buffer size is infinite. This is a reasonable assumption nowadays as smartphones can afford very large buffers where none of the packets will be discarded. 
%\textcolor{black}{In Section~\ref{sec:evaluation}, we show that  a buffer of $21$\:MB suffices for none of the packets to be dropped. This comes at the cost of a slight increase in the streaming latency.}  
%\textcolor{black}{Considering the case of finite-length buffers is beyond the scope of this paper, and will be part of our future work.} 

%\emph{Include some survey results on users' satisfaction with consistent QoE} 
%According to~\cite{Nam16}, variable play out rates are the second cause of users abandoning video streaming, resulting in up to $21$\% of users abandoning the video, with the main cause being the rebuffering (outage) events. On the other hand, Mux (mux.com) conveyed a survey with more than 1,000 US respondents~\cite{mux} about their experience with video streaming services. According to that survey, the low picture quality (low video resolution) was the most frustrating problem for $14.3$\% of the viewers, and $57$\% of them have abandoned a video in the past due to low video resolution. Hence, the importance of not only minimizing the video outage due to rebuffering, but also of providing a high video resolution in playing the video.       

Since the model is adjusted for \textcolor{black}{live} video streaming, there are always packets ``circulating in the air''. We assume that at time $t=0$, when streaming starts, there are no packets in the buffer, i.e., $Q(0)=0$. %which is reasonable.  
Packets ``start'' arriving at the buffer during the first frame \textcolor{black}{or later}, and during that time, %i.e., the first frame, 
there is no video being played out. The playout starts %at frame %$t=1$. 
%as soon as 
%\textcolor{black}{in the frame succeeding }
\textcolor{black}{after} the first packets \textcolor{black}{have arrived} at user's buffer. 
We assume that at the beginning of the frame the video playout is preceded by the arrival of the packets in the buffer. Hence, our ``point of interest'' is \emph{the beginning of every frame \textcolor{black}{just after the %batch 
arrival of the last packet from the previous frame} and just before the video resuming.} We denote with $Q_B(t)$ the number of packets in the buffer at \textcolor{black}{that} moment.

As data rate changes over time, the number of packets arriving in the queue \textcolor{black}{varies over frames}. %is different in different frames, in general. 
The buffer amortizes the throughput variability while providing a constant playout rate. Since the goal is to provide a %consistent 
\textcolor{black}{high} QoE to every user, we strive to have as high a constant playout rate %guaranteed 
as possible with a very small outage $\epsilon$.\footnote{\textcolor{black}{This means that the user will have the same playout rate while watching the entire video.}} %Assume that playout rate is $U$. 
If $U$ is the playout rate, the total number of packets being played out during a frame \textcolor{black}{(while the streaming lasts)} is
%Then, during a frame the total number of packets being played out is 
$S=\lfloor{\frac{U\Delta t}{\sigma}\rfloor}$. \textcolor{black}{We assume that the buffer size is larger than any reasonable number of packets that can be played out during a frame.}

%\textcolor{blue}{We consider the state of the buffer $Q(t)$ at the point after the arrival of all the packets from the previous frame and just before the packets have started to play out.}

As already mentioned, the outage refers to the amount of stalling in the video, which corresponds to the case when there are no packets to be played, i.e., the buffer becomes empty during a frame. This is equivalent to the ratio of frames during which there are fewer packets in the buffer than that can be played out. Consequently, we are interested in the stationary  probability that at the beginning of the frame there are 
fewer than $S$ packets in the buffer, in which case, based on the model, at some point during the upcoming frame there would be no packets, i.e., a stall in the video will occur. 

\textcolor{black}{As will become clear in the next section, besides the distribution of the state of the finite buffer, we will also need the distribution of the infinite-buffer state. To differentiate between the two, we use $q_{i,B}=\mathbb{P}(Q_B=i)$ to denote the stationary probability of finding $i$ packets in the queue of the finite-size buffer just before playout ``resumes'' at the beginning of the frame (and right after the packets from previous frame have arrived at the queue), whereas $q_{i}=\mathbb{P}(Q=i)$ does that for that infinite-size buffer.}

%Summarizing, if $q_{i}=\mathbb{P}(Q=i)$ denotes the stationary probability of finding $i$ packets in the queue just before playout \textcolor{black}{``resumes''} at the beginning of the frame (\textcolor{black}{and right after the packets from previous frame have arrived at the queue}), we are interested in finding $S$ for which it holds $\sum_{i=0}^{S-1}q_i\leq\epsilon$.

\textcolor{black}{Summarizing, we are interested in finding $S$ for which it holds $\sum_{i=0}^{S-1}q_{i,B}\leq\epsilon$ and $\delta\leq\delta_0$.}

Based on the above description, our system corresponds to a discrete-time \textcolor{black}{D[X]/D/1}~\cite{Bisdikian96} or GI/D/S~\cite{Mieghem06} queue \textcolor{black}{with finite buffers.} ~%\cite{Bruneel93_book},
 
%where   
%$GI$ denotes the general number of packets at the beginning of the frame that is independent of the frame, whereas $D$ denotes the deterministic service time of $\frac{\Delta t}{S}$ (the same number of packets in a frame if in the buffer there are at least $S$ packets). 
%\textcolor{black}{or even as a discrete-time GI/D/S queue ~\cite{Mieghem06}}.
%This can be considered also as a GI/D/S queue~\cite{Mieghem06} in which the service time is $\Delta t$, i.e., one frame, instead of $\frac{\Delta t}{S}$, but there are $S$ servers.  %Basically, it takes every packet the entire frame to be played out, but there are $K$ servers in parallel. 

The important question that arises is: \emph{What is the maximum value of $U$ that can be guaranteed to a user so that the outage probability is not greater than $\epsilon$ \textcolor{black}{and that the drop rate is not higher than $\delta_0$}?} 
In the next section, we provide the analysis that answers this question, i.e., that determines the maximum possible value of the number of packets $S$ that can be played out during a frame, resulting in the maximum possible \textcolor{black}{playout rate} (video resolution) that can be guaranteed to a mobile user for $1-\epsilon$ of the time, \textcolor{black}{while not losing more than $\delta_0\cdot100$\% of the packets}. 

%Before proceeding any further,  
Table~\ref{table:notation} summarizes the notation used throughout this paper. \textcolor{black}{In Section~\ref{sec:analysis}, to ease the presentation, we remove the reference to user $i$.} 
\begin{table}
\centering
\caption{Definitions and Notation}\label{table:notation}
%\vspace{-6pt}
\begin{tabular}{|l|l|} \hline
%\footnotesize
 $n$ & Number of users in the cell \\\hline 
 $R_i(t)$ & Per-block rate of user $i$ in frame $t$ \\\hline
 %AMC & Adaptive modulation and coding scheme\\\hline
 %$p_{R_i}(x)=\mathbb{P}\left(R_i=x\right)=p_{i,x}$ & PMF of user's $i$ per-block rate \\\hline
 $m$ & Number of levels in the MCS\\\hline
 %$l$ & Number of different video resolutions\\\hline
 $K$ & Total number of blocks\\ \hline
 $B$ & Size of the buffer\\ \hline
 $q_{i}$ & Probability of having $i$ packets in the infinite-size buffer\\\hline
$q_{i,B}$ & Probability of having $i$ packets in the finite-size buffer\\\hline 
$\epsilon$  & Outage probability\\\hline
$\delta$ & Packet drop rate \\ \hline
$C_i(t)$  & Data rate of user $i$ during frame $t$\\\hline
$A_i(t)$  & Number of packets arriving at user $i$ during frame $t$\\\hline
$A(z)$ & Probability generating function of the arrival process\\\hline
$Q(t)$ & Number of packets in the buffer in frame $t$\\\hline
$Q(z)$ & Probability generating function of the buffer state\\\hline
$U$ & Playout rate\\\hline
%$U_{min}$ & Minimum playout rate corresponding to $f_1$\\\hline
%$U_{max}$ & Maximum playout rate corresponding to $f_l$\\\hline
%$M$ & Buffer size\\\hline
$S$ & Number of packets that can be played out during a frame \\\hline
$\Delta t$ & Frame duration\\\hline
$\sigma$ & Packet size \\\hline
$Y_i$ & Ratio of frame (frame ratio) resources are allocated to user $i$\\\hline
\end{tabular}
%\normalsize
%\vspace{-10pt}
\end{table}

\textcolor{black}{\emph{Maintaining low playout delay:} As the backhaul link capacities are usually much higher than those of access network, we assume that the packets ``appear’’ instantly at the base station buffer. The buffer at the base station is much larger than the user’s buffer. Hence, it can be assumed to be infinite. %The latency on the user’s side is controlled by the size of the buffer. 
The playout delay, i.e., the latency component on the user's side is controlled by the size of its buffer. 
%To preserve the information freshness, 
To keep the information ``fresh'', 
i.e., in order to prevent a packet to exceed the maximum latency, it can be dropped from the BS buffer after a certain time. E.g., assume that the total latency allowed is $10$\:s. If the size of the buffer on the user’s side is $3$\:s, then as soon as a packet in BS buffer spends more than $7$\:s it is discarded. Nevertheless, this does not affect our analysis as we assume that the arrival process is generic. Therefore, we consider only the queueing process at the user’s side. }

\section{Playout Rate Analysis}
\label{sec:analysis}
In our analysis we first derive the results under the assumption that the buffer size at the user device is infinite, and then use those results to derive the analysis for the finite-buffer case. %This is followed by the analysis corresponding to finite-buffer size, where we use the results of the former.

\subsection{Infinite-size buffer}
\label{sec:analysis_infinite}
We solve the queue that corresponds to our system, where the arrival process \textcolor{black}{of number of packets} is $A(t)$.  
The probability generating function (PGF) for the arrival process is
%\vspace{-6pt}
%\footnotesize
\begin{equation}
A(z)=\sum_{i=0}^{\infty}a_i z^i, \quad |z|\leq 1,
\label{eq:generating_function}
\end{equation}
%\normalsize
where $a_i=\mathbb{P}(A=i)$. The PGF for the number of packets in the queue (including the one in service) is
%\vspace{-6pt}
%\footnotesize
\begin{equation}Q(z)=\sum_{i=0}^{\infty}q_i z^i, \quad |z|\leq 1.
\label{eq:Q_PGF}
\end{equation}
%\normalsize %where $q_i=\mathbb{P}(Q=i)$.
The evolution of the number of packets in the queue is
%\vspace{-6pt}
%\footnotesize
\begin{equation}
Q(t+1)=\max\{S, Q(t)\}-S+A(t),
\label{eq:queue}    
\end{equation}
%\normalsize
where $Q(0)=0$. For the steady-state regime, which is of our interest, we have 
$Q=\lim_{t \to \infty}Q(t).$ 
The sequence of the number of queued packets $Q$  forms a \textcolor{black}{stationary}  Markov chain \textcolor{black}{($A(t)$ is i.i.d.)}. %In this paper, 
%We are only concerned with the case \textcolor{black}{where}  $\mathbb{P}\left(A\geq S\right)>0$, since for $\mathbb{P}\left(A\geq S\right)=0$ the evolution of the queue is trivial. %Namely, then at our moments of interest, there would be never be any packets. 
%Besides the aforementioned condition, 
\textcolor{black}{This Markov chain is irreducible, positive recurrent, and aperiodic (which means there is a stationary distribution $q_i=\mathbb{P}[Q=i]$) 
if $\mathbb{P}[A\leq S-1]>0$, $\mathbb{P}[A\leq S]<1$, and $\rho=\frac{\mathbb{E}[A]}{S}<1$, which are compliant with our system.}    
%In order for this Markov chain to be ergodic, i.e., aperiodic and irreducible, another condition must hold, that of stability expressed as
%$\mathbb{E}[A]<S$, or equivalently $\rho=\frac{\mathbb{E}[A]}{S}<1$.

In equilibrium, Eq.(\ref{eq:queue}) becomes $Q=T-S+A$, where $T=\max\{S, Q\}$. \textcolor{black}{As $A(t)$ is independent of $Q(t)$ and $S$ is constant, the PGF of the queue can be written as} 
%Hence, we can write the PGF of the queue as  
%\vspace{-6pt}
%\footnotesize
\begin{equation}
Q(z)=\mathbb{E}\left[z^Q\right]=\mathbb{E}\left[z^{T-S+A}\right]=\mathbb{E}\left[z^T\right]\mathbb{E}\left[z^A\right]\mathbb{E}\left[z^{-S}\right].
\end{equation}
%\normalsize
Since $\mathbb{E}[z^{-S}]=z^{-S},$  $\mathbb{E}[z^A]=A(z)$, and $\mathbb{E}[z^T]=T(z)$, we have
%\vspace{-6pt}
%\footnotesize
\begin{equation}
Q(z)=z^{-S}A(z)T(z).
\label{eq:q_z_0}
\end{equation}
%\normalsize
The expression for $T(z)$ can be further transformed into
%\vspace{-6pt}
%\footnotesize
$$T(z)=\sum_{i=0}^{S-1}q_iz^S+\sum_{i=S}^{\infty}q_iz^i=z^S\sum_{i=0}^{S-1}q_i+\sum_{i=0}^{\infty}q_iz^i-\sum_{i=0}^{S-1}q_iz^i,$$%\normalsize
or, equivalently into
%\vspace{-9pt}
%\footnotesize
$$T(z)=Q(z)+\sum_{i=0}^{S-1}q_i(z^S-z^i)=Q(z)+\sum_{i=0}^{S-1}q_iz^i(z^{S-i}-1).$$ %\normalsize
The expression $z^{S-i}-1$ in the above equation transforms into
%\vspace{-9pt}
%\footnotesize
$$z^{S-i}-1=(z-1)(z^{S-i-1}+z^{S-i-2}+\ldots+1)=(z-1)\sum_{l=0}^{S-i-1}z^{l}.$$ %\normalsize
Hence, for $T(z)$ we obtain
%\vspace{-6pt}
%\footnotesize
$$T(z)=Q(z)+(z-1)\sum_{i=0}^{S-1}q_iz^i\sum_{l=0}^{S-i-1}z^{l},$$%\normalsize 
which is equivalent to
%\vspace{-6pt}
%\footnotesize
$$T(z)=Q(z)+(z-1)\sum_{i=0}^{S-1}q_i\sum_{l=i}^{S-i-1}z^{i+l}$$ %\normalsize
because $\sum_{l=0}^{S-i-1}z^{i+l}=\sum_{l=i}^{S-1}z^{l}.$ Denoting $N(z)=\sum_{i=0}^{S-1}q_i\sum_{l=i}^{S-1}z^l,$ we get
%\vspace{-6pt}
%\footnotesize
\begin{equation}
T(z)=Q(z)+(z-1)N(z).
\label{eq:Y}
\end{equation}
%\normalsize
Finally, replacing Eq.(\ref{eq:Y}) into Eq.(\ref{eq:q_z_0}), and rearranging we obtain 
%\vspace{-9pt}
%\footnotesize
\begin{equation}
Q(z)=\frac{(z-1)N(z)A(z)}{z^S-A(z)}.
\label{eq:queue_distribution}
\end{equation}
%\normalsize
%Since 
$Q(z)$ is an analytic function, \textcolor{black}{i.e. differentiable},  everywhere inside \textcolor{black}{and on} the unit circle. %As $Q(1)=1$, we have
%For $z=1$, from Eq.(\ref{eq:queue_distribution}) we have
\textcolor{black}{Eq.(\ref{eq:queue_distribution}) for $z=1$ yields}
%\vspace{-6pt}
%\footnotesize
\begin{equation}
Q(1)=\frac{\lim_{z\to 1} (z-1)N(z)A(z)}{\lim_{z\to 1} \left(z^S-A(z)\right)}=\frac{\lim_{z\to 1}\frac{d}{dz}\left((z-1)N(z)A(z)\right)}{\lim_{z\to 1} \frac{d}{dz}\left(z^S-A(z)\right)}.
\label{eq:Q1_new}
\end{equation}
%\normalsize
The RHS in the previous equation is obtained after applying L'Hopital's rule to the LHS (a ratio of the form $\frac{0}{0}$). Applying some basic calculus to the numerator, we get
%\vspace{-9pt}
%\footnotesize
$$\lim_{z\to 1}\frac{d}{dz}\left((z-1)N(z)A(z)\right)=\lim_{z\to 1}\frac{d}{dz}\left(zN(z)A(z)-N(z)A(z)\right)=$$
$$\lim_{z\to 1}\left(\frac{d}{dz}\left(zN(z)\right)A(z)+zN(z)A^{'}(z)-N^{'}(z)A(z)-N(z)A^{'}(z)\right)=$$ %\normalsize
%This leads to
%\vspace{-6pt}
%\footnotesize
\begin{equation*}
\lim_{z\to 1}\left(N(z)A(z)+zN^{'}(z)A(z)+zN(z)A^{'}(z)-N^{'}(z)A(z)-N(z)A^{'}(z)\right),
\end{equation*}
%\normalsize
resulting in the numerator being simply $N(1)A(1)$. Since $A(1)=1$, for the numerator we have
%\vspace{-6pt}
%\footnotesize
\begin{equation}
\lim_{z\to 1}\frac{d}{dz}\left((z-1)N(z)A(z)\right)=N(1).
\label{eq:numerator}
\end{equation}
%\normalsize
When it comes to the denominator of the expression for $Q(1)$, %we have 
%\vspace{-9pt}
%\footnotesize
$$\lim_{z\to 1} \frac{d}{dz}\left(z^S-A(z)\right)=\lim_{z\to 1}\left(Sz^{S-1}-A^{'}(z)\right)=S-A^{'}(1).$$
%\normalsize
Note that $A^{'}(1)=\sum_{i}ia_i=\mathbb{E}[A]$, leading to the denominator being simply 
%\vspace{-6pt}
%\footnotesize
\begin{equation}
\lim_{z\to 1} \frac{d}{dz}\left(z^S-A(z)\right)=S-\mathbb{E}[A].    
\label{eq:denominator}
\end{equation}
%\normalsize
%As already mentioned, 
From \textcolor{black}{Eq.(\ref{eq:Q_PGF}),} $Q(1)=1$, %then  from 
which combined with 
Eq.(\ref{eq:numerator}) and Eq.(\ref{eq:denominator}) into Eq.(\ref{eq:Q1_new}) yields %\vspace{-6pt}
%\footnotesize
\begin{equation}
    N(1)=S-\mathbb{E}[A]=S\left(1-\frac{\mathbb{E}[A]}{S}\right)=S(1-\rho). 
\label{eq:N_1}    
\end{equation}
%\normalsize
%where $\rho=\frac{\mathbb{E}[A]}{S}$ is the utilization ratio of the system.
%Due to the fact that $Q(z)$ is an analytic function 
%in $|z|\leq 1$, its poles, i.e., the zeros of its denominator, must be the same as its zeros, i.e., the zeros of its numerator, 
%\textcolor{blue}{As $Q(z)$ is analytic in $|z|\leq 1$, its poles, i.e., the zeros of $z^S-A(z)$, must be the same as its zeros, i.e., the zeros of $(z-1)N(z)$.}  
\textcolor{black}{As $Q(z)$ is analytic in $|z|\leq 1$, the zeros of $z^S-A(z)$ must be the same as the zeros of $(z-1)N(z)$.}
%\textcolor{blue}{implying } 
\begin{lemma}
There are exactly $S$ \textcolor{black}{roots} of $z^S-A(z)$ satisfying $|z|\leq 1$, of which exactly $S-1$ in $|z|< 1$.
\label{lemma}
\end{lemma}
%\begin{proof}
Due to space limitations \textcolor{black}{and because of the fact that it is not of further technical interest}, we omit the proof of this lemma. It can be found in~\cite{Mieghem06}. %~\cite{Takacs62}.
%\end{proof}

We use the $S-1$ roots 
%According to Lemma~\ref{lemma} there are a total of $K-1$ roots 
$z_1, \ldots, z_{S-1}$ within $|z|< 1$  and one other root for which it holds $z_S=1$ to form a system  %We use these values to form a system 
of $S$ linear equations in the unknowns $q_0, \ldots, q_{S-1}$ from the expression for $N(z)$. 
%\footnote{Note: This is a solution for a given $K$.}
Namely, 
%Based on the previous discussion, 
it holds that $N(z_i)=0, \ \forall{i\in\{1,\ldots,S-1\}},$ and $N(z_S)=\sum_{i=0}^{S-1}q_i(S-i)=S-\mathbb{E}[A]$, where the latter is given in  Eq.(\ref{eq:N_1}). Applying Cramer's rule, we solve this system of equations to obtain
%\vspace{-6pt}
%\footnotesize
\begin{equation}
    q_i=\frac{D_{q_i}}{D}, \quad \forall{i\in\{0,\ldots,S-1\}},
\end{equation}
%\normalsize
where %\vspace{-6pt} %\footnotesize
$$D=\begin{vmatrix}
\sum_{l=0}^{S-1}z_{1}^{l} & \sum_{l=1}^{S-1}z_{1}^{l} & \ldots & z_1^{S-1}\\ 
\sum_{l=0}^{S-1}z_{2}^{l} & \sum_{l=1}^{S-1}z_{2}^{l} & \ldots & z_2^{S-1}\\ 
\ldots & \ldots & \ldots & \ldots\\
\sum_{l=0}^{S-1}z_{S-1}^{l} & \sum_{l=1}^{S-1}z_{S-1}^{l} & \ldots & z_{S-1}^{S-1}\\ 
S & S-1 & \ldots & 1\\
\end{vmatrix},$$ %\normalsize
and 
%$$D_{q_i}=\begin{vmatrix}
$D_{q_i}$ is the same as the determinant $D$ except for the $i$th column,  which is %\footnotesize
$[\underbrace{0 \ldots 0}_{S-1} \quad S-\mathbb{E}[A]]^{tr}$. %\normalsize. %where $0$ appears $S-1$ times. %(each corresponding to a zero $|z|<1$).
%\end{vmatrix}
%$$

%Note that these determinants never produce complex numbers. The reason is that some of the roots are real numbers whereas some others may be complex-conjugate pairs. 
%However, due to the structure of these determinants the imaginary parts will cancel out. 
Due to the structure of these determinants, we have only real numbers in $(0,1)$ for $q_i$. For a given $S$, we obtain as solutions $q_i, \forall{i}=0,\ldots,S-1$, which is what we need. %Let's look at this in a simple exam. But, first note that for complex pairs it holds that $\bar{z^{l}}=\bar{z}^l$ \cite{}. If we have the determinant

\subsection{Finite-size buffer}
\label{sec:analysis-finite}
%Having an infinite size buffer can be unrealistic, especially in the context of real-time video streaming where the goal is to minimize the latency, which inevitably imposes constrained buffers~\cite{XX}. Therefore, 
We proceed with the analysis for the finite-size buffer. As already mentioned in Section~\ref{sec:model}, %consider the analysis for finite-buffer size next. As will be seen, we need the results for infinite-size buffers. To this end, we consider buffers of size $B$. 
we assume that $B>S$. We denote the state of the finite-size buffer with $Q_{B}(t)$.  

The evolution of the number of packets in the buffer evolves according to
\begin{equation}
Q_{B}(t+1)=\min\{B,\max\{S,Q_{B}(t)\}-S+A(t)\}.
\label{eq:finite_buffer}
\end{equation}
Following the same reasoning as with infinite-size buffers, it can be shown that the process Eq.(\ref{eq:finite_buffer}) is a Markov chain, and that it is ergodic. Its transition probability matrix is 
\begin{equation}
\Psi=\left[\psi_{i,j}, 0\leq i,j\leq B\right].
\end{equation}
The individual transition probabilities depend on the state of the system. So, for the elements of the transition probability matrix $\Psi$ we have:  
$$\psi_{i,j}=a_j, \quad i\in\{0,\ldots,S\}, \ 0\leq j\leq B-1,$$
$$\psi_{i,j}=0, \quad i\in\{S+1,\ldots,B\}, \ 0\leq j\leq i-S-1,$$
$$\psi_{i,j}=a_{j-i+S}, \quad i\in\{S+1,\ldots,B\}, \ i-S\leq j\leq B-1,$$
$$\psi_{i,B}=1-\sum_{k=0}^{B-1}\psi_{i,k}, \quad 0\leq i\leq B, \ j=B,$$
The state of the buffer in equilibrium is denoted as
$$Q_{B}=\lim_{t\to\infty}Q_{B}(t).$$
The probability of having $i$ packets in the buffer is
$$q_{i,B}=\mathbb{P}\left[Q_{B}=i\right], \quad 0\leq i\leq B.$$
Written in compact form, the probability mass function of the state of the buffer is 
$$q_{B}=\left[q_{0,B},\ldots,q_{B,B}\right].$$
The steady-state probabilities of the buffer state $q_{i,B}, \ i\in\{0,\ldots,B\}$ are determined by solving the system of equations
\begin{equation}
q_B\Psi=q_B, \quad \sum_{i=0}^{B}q_{i,B}=1.
\end{equation}
The average number of packets being transmitted per frame is
\begin{equation}
\beta_B=\sum_{i=0}^{S-1}i q_{i,B}+S\mathbb{P}\left[Q_B\geq S\right],    
\end{equation}
where the first RHS term corresponds to the case when there are fewer packets in the buffer than the packets that can be played out in a frame. The second term denotes the instances when there are more packets in the buffer than packets that are played out per frame. Equivalently, we have
\begin{equation}
\beta_B=\sum_{i=0}^{S-1}i q_{i,B}+S-S\mathbb{P}\left[Q_B< S\right].    
\end{equation}
%We have the equivalent expression
This is equivalent to  
%\begin{equation}
%\beta_B=S+\sum_{i=0}^{S-1}i q_{i,B}-S\sum_{i=0}^{S-1}q_{i,B},  
%\end{equation}
%and
\begin{equation}
\beta_B=S-\sum_{i=0}^{S-1}(S-i) q_{i,B}.  
\end{equation}
Given that we are dealing with an ergodic process, the probability of a packet being dropped is 
\begin{equation}
\delta=1-\frac{\beta_B}{\mathbb{E}[A]},    
\end{equation}
where $\mathbb{E}[A]$ is, as mentioned in Section~\ref{sec:analysis_infinite}, the average number of packet arrivals per frame. Further, we have
\begin{equation}
\delta=\frac{\sum_{i=0}^{S-1}(S-i)q_{i,B}-N(1)}{S-N(1)}.    
\end{equation}
Replacing $N(1)$ from the previous section in the previous equation, we obtain
\begin{equation}
\delta=\frac{\sum_{i=0}^{S-1}(S-i)q_{i,B}-\sum_{i=0}^{S-1}(S-i)q_i}{S-\sum_{i=0}^{S-1}(S-i)q_i}.    
\end{equation}
After dividing both the numerator and denominator by $S$, we have
\begin{equation}
\delta=\frac{\sum_{i=0}^{S-1}(1-\frac{i}{S})\left(q_{i,B}-q_{i}\right)}{1-\sum_{i=0}^{S-1}(1-\frac{i}{S})q_{i}}.    
\end{equation}
Finally, we have the following result:
\begin{result}
The maximum playout rate with maximum  outage $\epsilon$ and maximum allowed packet drop rate $\delta_0$ for a user whose data rate $C(t)$ \textcolor{black}{corresponds to arrival process $A(t)$} is
%\vspace{-6pt}
%\footnotesize
\begin{equation}
U=\max\left\{\frac{S\sigma}{\Delta t}\middle| \sum_{i=0}^{S-1}q_{i,B}\leq\epsilon, \ \frac{\sum_{i=0}^{S-1}(1-\frac{i}{S})\left(q_{i,B}-q_{i}\right)}{1-\sum_{i=0}^{S-1}(1-\frac{i}{S})q_{i}}\leq \delta_0\right\}.   
\label{eq:U_c_finite}
\end{equation}
%\normalsize
%where $S=\lfloor{\frac{U\Delta t}{\delta}\rfloor}$.
\end{result}

\textcolor{black}{The result can be obtained numerically. Note that as $U$, and hence $S$ increases, the probability mass function of the number of packets in the queue shifts towards lower values, i.e., for higher playout rates there is a higher probability that fewer packets will be encountered in the queue compared to the lower playout rates when there is a tendency of having more packets queued. Consequently, since there is a constraint of having less than $S$ packets, the playout rate can only be increased up to the point where it doesn't violate the constraint $\sum_{i=0}^{S-1}q_{i,B}\leq\epsilon$, and that is the maximum achieved playout rate.} 

\textcolor{black}{Having the buffer size expressed in seconds instead of data units can be captured by our model too. Namely, if the buffer size is $L$ seconds, then $B=\lfloor{\frac{U_pL}{\sigma}\rfloor}$. Hence, $B$ is replaced by $\lfloor{\frac{U_pL}{\sigma}\rfloor}$ throughout the analysis in Section~\ref{sec:analysis-finite}. Since Result 1 is obtained numerically, having $U_p$ on both sides of Eq.(\ref{eq:U_c_finite}) is not an issue.}
\textcolor{black}{The value of $S$ remains unchanged across all the frames, i.e., the playout rate $U_p$ is fixed over time (it is not adjustable).}

\textcolor{black}{This model is very important as it also enables solving the inverse problem, that of buffer dimensioning, i.e., determining the value of $B$, given $U_p$, $\epsilon$, and $\delta$.}

\textcolor{black}{If the packet size is a function of the video resolution at which it is going to be played out, our model can capture that intricacy too. Namely, if the packet size is proportional to the resolution, we denote it as $\sigma_f$, then the playout rate has to be proportional to the video resolution, $U_f$. We assume w.l.o.g. that the proportionality ratio on the resolution is the same for the packet size and the playout rate. Hence, the time it takes to playout a packet,  $t_f=\frac{\sigma_f}{U_f}=const$, is constant and independent of the video resolution. This results in a fixed number of packets that can be played out per frame for any resolution. However, the arrival process is a function of the packet size. The higher the resolution, the fewer packets arriving to the user’s buffer. Therefore, this is a special case of our model, and can be solved by determining the video resolution (from the arrival process) such that the requirements on $\epsilon$ and $\delta$ are not violated.}

%\textcolor{blue}{Approach 2: Assume that only the packet size is a function of the video resolution, but not the playout rate. Then, the number of packets that can be served per frame is a function of the video resolution. The same holds for the arrival process. Plugging these constraints in our model, we can compute the size of the packet (and hence the video resolution) and the playout rate from our model, such that the constraints on $\epsilon$ and $\delta$ are satisfied.}

\section{%Performance optimization 
Optimal Resource Allocation}
\label{sec:optimization}
We solve three problems in this section.  First, we consider the problem of providing the \textcolor{black}{same}  %highest possible 
\textcolor{black}{achievable} video resolution to everyone. %This is followed 
Then, the problem of maximizing the number of users that experience a given resolution is solved, \textcolor{black}{while providing a minimum guaranteed resolution to the others.} %Finally, we consider the case of allocating resources to provide proportional fairness in terms of the playout rate.
Finally, we consider the case of two classes of users. 

%we consider the case of maximizing the number of users that will experience the maximum possible video resolution with a given outage. This is followed by the problem in which the objective is to provide the same highest possible resolution to every user in the cell. Finally, in the third problem we consider the case of providing proportional fairness in terms of the playout rate over all users in the cell.

\subsection{Same experience to everyone}
\label{subsub:same}
In this problem, \textcolor{black}{the goal is to provide 
the same playout rate to all the users with the same outage probability \textcolor{black}{and drop rate}.} 
%the same QoE to all the users, expressed as the playout rate (video resolution) the video could be played at with the same outage probability. 
This leads to the natural question of \emph{how to allocate the resources to these users to accomplish this objective, given that they have different channel characteristics}? 

%We consider two cases for this scenario. In the first one, there is no consistent data rate provided, but rather the resources are allocated in such a way that there is a constant playout rate provided with the same outage $\epsilon$. In the second case, first the consistent rate is provided and then the unused resources are reallocated in such a way that all users will experience the same maximum possible playout rate $U_{max}$. Finally, we will compare the performance of these two approaches in Section~\ref{sec:evaluation}. 

If $Y_i(t)$ denotes the frame ratio all the $K$ blocks are allocated to user $i$ in frame $t$, where her per-block rate is $R_i(t)$, the data rate that user receives in frame $t$ is 
%\vspace{-6pt}
%\footnotesize
\begin{equation}
C_i(t)=KY_i(t)R_i(t).    
\label{eq:C} 
\end{equation}
%\normalsize
The problem reduces to determining the allocations $Y_i$, for which it holds
%\vspace{-6pt}
%\footnotesize
\begin{equation}
U_1(\epsilon,\delta_0)=\ldots=U_n(\epsilon,\delta_0).    
\end{equation}
%\normalsize
%In this problem we have the following optimization problem
%\begin{alignat}{2}
% \underset{Y_1,\ldots,Y_n}{\text{max}} \quad
% &U_1(\epsilon)=\ldots=U_n(\epsilon)\label{eq:objective} \\
% \text{s.t.}\\
% &\sum_{i=1}^{n}Y_i\leq 1,\label{eq:kushti_01}\\
% &  Y_i\geq 0, \quad  \forall i=1, \ldots,  n\label{eq:kushti_03}.
%\end{alignat}

Given the different channel characteristics over time of the different users, %in order 
\textcolor{black}{a way} to provide the same highest achievable playout rate %a way to do that 
is by ensuring that all users receive the same data rate (among themselves) in the frame. Of course, this data rate will be different in different frames. %but still the same data rate among all the users within the frame. 
If the per-channel rates of the $n$ users in frame $t$ are $R_1(t),\ldots,R_n(t)$, every user will receive the same rate if the base station resources are allocated inversely proportionally to user's per-block rate, i.e., users with good channel characteristics will receive fewer resources than users with bad channel characteristics. Hence, user $i$ will receive \textcolor{black}{the following frame ratio} of network resources
%\vspace{-6pt}
%\footnotesize
\begin{equation}
Y_i(t)=\frac{\frac{1}{R_i(t)}}{\sum_{j=1}^{n}\frac{1}{R_j(t)}}. 
\label{eq:Yi_equal}
\end{equation}
%\normalsize
The data rate of user $i$ in frame $t$ with this policy is obtained substituting Eq.(\ref{eq:Yi_equal}) into Eq.(\ref{eq:C}), and is 
%\vspace{-6pt}
%\footnotesize
\begin{equation}
C_i(t)=C(t)=\frac{K}{\sum_{j=1}^{n}\frac{1}{R_j(t)}}.    
\label{eq:rate_same}
\end{equation}
%\normalsize
%In the second case, $C_{c,i}(t)=C_c(t)=\frac{KY(t)}{\sum_{j=1}^{n}\frac{1}{R_j(t)}}$
As can be seen from Eq.(\ref{eq:rate_same}), all the users will receive the same data rate in the frame. 
\subsubsection{The distribution of $C(t)$} 
Next, we need to determine the \textcolor{black}{probability mass function} (PMF) of $C$. It is defined as
%\vspace{-6pt}
%\footnotesize
\begin{equation}
p_{C}(x)=\mathbb{P}(C=x)=\mathbb{P}\left(\frac{K}{\sum_{j=1}^{n}\frac{1}{R_j}}=x\right)=\mathbb{P}\left(\sum_{j=1}^{n}\frac{1}{R_j}=\frac{K}{x}\right).  
\label{eq:convolution_0}
\end{equation}
%\normalsize
This further leads to
%\vspace{-6pt}
%\footnotesize
\begin{equation}
\mathbb{P}\left(\sum_{j=1}^{n}\frac{1}{R_j}=\frac{K}{x}\right)=\mathbb{P}\left(\frac{1}{R_1}=y\right)*\ldots * \mathbb{P}\left(\frac{1}{R_n}=y\right) _{y=\frac{K}{x}},
\label{eq:convolution}
\end{equation}
%\normalsize
where * denotes the convolution operation. The RHS of Eq.(\ref{eq:convolution}) is equivalent to
%\vspace{-6pt}
%\footnotesize
\begin{equation}
\mathbb{P}\left(R_1=\frac{1}{y}\right)*\ldots * \mathbb{P}\left(R_n=\frac{1}{y}\right) _{y=\frac{K}{x}}.
\label{eq:convolution_2}
\end{equation}
%\normalsize
Substituting Eq.(\ref{eq:convolution_2}) into Eq.(\ref{eq:convolution_0}), we obtain
%\vspace{-6pt}
%\footnotesize
\begin{equation}
p_C(x)=p_{R_1}\left(\frac{1}{y}\right)*\ldots*p_{R_n}\left(\frac{1}{y}\right)_{y=\frac{K}{x}}.
\label{eq:convolution_3}
\end{equation}
%\normalsize
In Eq.(\ref{eq:convolution_3}), $p_{R_i}(x)$ is the PMF of user's $i$ per-block rate, which is expressed as \textcolor{black}{a sum of weighted Dirac delta functions $\delta(x)$:}
%\vspace{-9pt}
%\footnotesize
\begin{equation}
p_{R_i}\left(\frac{1}{y}\right)=\sum_{k_i=1}^{m}p_{R_i}(r_{k_i})\cdot\delta\left(y-\frac{1}{r_{k_i}}\right).
\label{eq:Ri}
\end{equation}
%\normalsize
As the convolution of a signal with a shifted Dirac delta function is just the shifted signal itself~\cite{Oppenheim}, after some calculus operations on Eqs.(\ref{eq:convolution_3}) and (\ref{eq:Ri}), we obtain
%\vspace{-6pt}
%\footnotesize
\begin{equation}
p_C(x)=\sum_{k_{1}=1}^{m}\dots\sum_{k_{n}=1}^{m}p_{R_1}(r_{k_1})
\ldots p_{R_n}(r_{k_n}) \delta\left(\frac{K}{x}-\frac{1}{r_{k_{1}}}-\ldots-\frac{1}{r_{k_{n}}}\right).    
\label{eq:p_c}
\end{equation}
%\normalsize
Further, the number of packets that arrive at the buffer of any user, following this policy, is 
%\vspace{-6pt}
%\footnotesize
\begin{equation}
A(t)=\frac{C(t)\Delta t}{\sigma}.    
\end{equation}
%\normalsize
The PMF of the arrival process $A(t)$ is
%\vspace{-6pt}
%\footnotesize
\begin{equation}
p_A(i)=\mathbb{P}\left(\frac{C\Delta t}{\sigma}=i\right)=\mathbb{P}\left(C=\frac{\sigma}{\Delta t}i\right)=p_C\left(\frac{\sigma}{\Delta t}i\right).  
\label{eq:p_A_0}
\end{equation}
%\normalsize
After replacing Eq.(\ref{eq:p_c}) into Eq.(\ref{eq:p_A_0}), we get
%\vspace{-6pt}
%\footnotesize
\begin{equation}
p_A(i)=\sum_{k_{1}=1}^{m}\hspace{-3pt}\dots\hspace{-3pt}\sum_{k_{n}=1}^{m}p_{R_1}(r_{k_1})
\ldots p_{R_n}(r_{k_n}) \delta\left(\frac{K\Delta t}{\sigma i}-\frac{1}{r_{k_{1}}}-\ldots-\frac{1}{r_{k_{n}}}\right).  \label{eq:arrival_process} 
\end{equation}
%\normalsize
The PGF of the arrival process $A(z)$ can be determined by replacing $p_A(i)=a_i$ into Eq.(\ref{eq:generating_function}).  

Now that we have completely characterized the input process, using Eq.(\ref{eq:U_c_finite}) we can determine the maximum number of packets $S$ that can be processed in a 5G frame, such that \textcolor{black}{the outage and drop rate constraints are not violated}. From this, we obtain:
\begin{result}
The highest possible playout rate with maximum outage $\epsilon$ \textcolor{black}{and maximum drop rate $\delta_0$} that can be guaranteed to all the users in the cell  is given by Eq.(\ref{eq:U_c_finite}).
%\begin{equation}
% U=\frac{S\sigma}{\Delta t}.
%\end{equation}
\end{result}
%and the corresponding video resolution for all the users in the cell.

%\textcolor{red}{\subsubsection{Providing a constant data rate} If the constant data rate $U_c$ is provided to every user %for $1-\epsilon$ of the time 
%in line with \cite{Fidan_TMC}, the amount of frame ratio used is $\sum_{j=1}^{n}\frac{U_c}{KR_j(t)}$, leaving as unused resources the frame ratio of $Y(t)=1-\frac{U_c}{K}\sum_{j=1}^{n}\frac{1}{R_j(t)}$. In order to provide the same data rate, these unused resources need to be allocated  \textcolor{black}{inversely proportionally} to user's per-block rate, similar to Eq.(\ref{eq:Yi_equal}). The total data rate after reallocation is then
%\vspace{-6pt}
%\footnotesize
%\begin{equation}
%C_c(t)=U_c+\frac{KY(t)}{\sum_{j=1}^{n}\frac{1}{R_j(t)}}=\frac{K}{\sum_{j=1}^{n}\frac{1}{R_j(t)}},    
%\end{equation}
%\normalsize
%which is the same data rate as in our case (Eq.(\ref{eq:rate_same})). \textcolor{black}{Hence, providing a constant data rate and reallocating unused resources~\cite{Fidan_TMC} performs no better than our approach.}}  %Hence, \textcolor{black}{in this scenario}, there is no need to provide a \textcolor{black}{constant}  data rate for real-time video streaming, as the same performance can be \textcolor{black}{achieved} using our approach. 

%and the unused resources are reallocated to the same users, such that everyone gets the same rate, the input data rate in $1-\epsilon$ of the time is
%\begin{equation}
%C_{c,i}=U_c+KR_i\left(1-\frac{U_c}{K}\sum_{j=1}^{n}\frac{1}{R_j}\right)
%\end{equation}

\subsection{Maximum resolution to as many users as possible}
\label{subsec:ultimate}
In the second problem,  %to consider is the one in which 
the operator is interested in providing \textcolor{black}{a given (high) resolution, which may or many not be the ultimate video experience,} to as many users as possible, while ensuring that the rest of them will receive a minimum guaranteed QoE with the same $\epsilon$ and \textcolor{black}{$\delta_0$}. So, the idea is to \textcolor{black}{determine the} amount of resources \textcolor{black}{needed} to provide the guaranteed playout rate, and then to allocate the rest \textcolor{black}{of the resources} to maximize the number of users with the maximum resolution (\textcolor{black}{from now on}). %But, before that we propose an approximation.

%\subsubsection{High-utilization approximation} %As we saw, determining the maximum number of users that can receive the ultimate user experience is numerically cumbersome. 

%Instead, in the following we will propose an approximation that relies on the aforementioned fact about the smartphone buffers operating in the high-utilization regime. Assuming that the utilization level $\rho \to 1$, we get that $\mathbb{E}[C_i]=K\left(1-\sum_{i=1}^{n}Y_{i,min}\right)Y_i\mathbb{E}[R_i]\approx \bar{U}$, leading to

\subsubsection{Providing the minimum playout rate}
 Let $U_{min}$ be the minimum playout rate \textcolor{black}{to be guaranteed to every user}. %run a video with the worst quality (lowest resolution). 
 The data rate user $i$ receives is $C_i=KY_iR_i$. Given that the output is $U_{min}$, \emph{what is the value of $Y_i$ that provides that playout rate}? Determining this value is possible only by using the \emph{trial-and-error} method for different values of $Y_i$ until we find the one that as the output of our queueing system provides $U_{min}$. This is cumbersome. Instead, we use an approximation,  
 \textcolor{black}{for which as the departing point of the derivation we consider the infinite-size buffer.} 
 %Namely, 
 \textcolor{black}{In an infinite-size buffer,} 
 since the goal is to have a very low $\epsilon$, we assume that the smartphone queue operates in the high-utilization regime with $\rho \to 1$, which results in 
 $\mathbb{E}[KY_iR_i]\approx U_{min}$. %yielding $\mathbb{E}[Y_iR_i]=\frac{U_{min}}{K}, \forall i$. 

\textcolor{black}{In the case of finite-size buffer, as up to $\delta_0$ ratio of the packets are lost, the approximation is modified to $(1-\delta_0)\mathbb{E}\left[KY_{i}R_i\right]\approx U_{min}$}.
 The previous equation entails that $Y_i$ and $R_i$ can't be directly proportional. %\textcolor{blue}{as a user with better $R$ will have a higher $Y$, leading to a higher $\mathbb{E}[Y_iR_i]$ than a user with worse $R$.}
 %\textcolor{blue}{as a user with better $R$ will have a higher $\mathbb{E}[Y_iR_i]$ than a user with worse $R$.}
 Therefore, there are only two possible cases. Either $Y_i$ and $R_i$ are uncorrelated, or they are inversely proportional. In the first case, $Y_i$ is static, so we have  
%\vspace{-6pt}
%\footnotesize
\begin{equation}
Y_{i,min}=\frac{U_{min}}{(1-\delta_0)K\mathbb{E}[R_i]}. 
\label{eq:Y_min}
\end{equation}
%\normalsize
 
%\textcolor{black}{Let us have a close look at $\mathbb{E}[Y_iR_i]=\frac{U_{min}}{K}$. 
\textcolor{black}{In the second case, %it holds that \textcolor{red}{Assume first that there exists an inverse proportionality in the dynamic policy between $Y_i$ and $R_i$, i.e., 
$\text{Cov}(Y_i,R_i)<0$. Then, from the definition of Covariance, we have $\text{Cov}(Y_i,R_i)=\mathbb{E}[Y_iR_i]-\mathbb{E}[Y_i]\mathbb{E}[R_i]<0$, resulting in $\mathbb{E}[Y_{i}R_{i}]<\mathbb{E}[Y_{i}]\mathbb{E}[R_i]$, and  $\mathbb{E}[Y_{i}]>\frac{\mathbb{E}[Y_{i}R_{i}]}{\mathbb{E}[R_i]}=\frac{U_{min}}{(1-\delta_0)K\mathbb{E}[R_i]}=Y_{i,min}$. This shows that on average more resources are needed to maintain the  minimum playout rate when there is a negative correlation between the resources allocated to a user and her per-block rate.}  

%\textcolor{red}{When there is no correlation between $Y_i$ and $R_i$ we get the static case, where $Y_{i,min}$ is determined from Eq.(\ref{eq:Y_min}).}

\textcolor{black}{This result is of practical importance, as (\ref{eq:Y_min}) is a static policy meaning that the frame ratio is determined at the beginning and does not change over time for any user.} %, i.e., it is not a function of the corresponding per-block rate.}

%\textcolor{red}{While (\ref{eq:Y_min}) is a static policy because $Y_{i,min}$ is determined at the beginning and does not change over time for any user, i.e., it is not a function of the corresponding per-block rate, as we will see in the sequel, there is no guarantee that we could do better (spend less resources) with any dynamic policy.} 

In case the number of users in the cell is high and/or their channel qualities are poor, it may happen that $\sum_{i=1}^{n}Y_{i,min}>1$. Then, the network resources are not sufficient to provide even the minimum video quality. An \emph{admission control} policy must then be used to reduce the number of users that can be guaranteed a minimum playout rate with outage $\epsilon$ \textcolor{black}{and maximum drop rate of $\delta_0$}.

\subsubsection{Allocating the unused resources}
%The RHS of constraint (\ref{eq:kushti_01}) denotes the amount of network resources left unused after providing the minimum video resolution to every user in the cell.
After providing the minimum playout rate, the amount of resources left unused is $Y=1-\sum_{j=1}^{n}Y_{j,min}=1-\frac{U_{min}}{(1-\delta)K}\sum_{j=1}^{n}\frac{1}{\mathbb{E}\left[R_i\right]}$. %$Y=1-\sum_{j=1}^{n}Y_{j,min}=1-\frac{U_{min}}{K}\sum_{j=1}^{n}\frac{1}{\mathbb{E}[R_j]}$. 
The amount of extra resources needed \textcolor{black}{for} user $i$ to obtain the maximum playout rate is
%\vspace{-6pt}
%\footnotesize
%\begin{equation}
%\Delta Y_i=\frac{U_{max}-U_{min}}{K\left(1-\frac{U_{min}}{K}\sum_{j=1}^{n}\frac{1}{\mathbb{E}[R_j]}\right)\mathbb{E}[R_i]}. 
%\Delta Y_i=\frac{U_{max}-U_{min}}{K\mathbb{E}[R_i]}.
%\label{eq:delta_y}
%\end{equation}
%\normalsize
\begin{equation}
\Delta Y_i=\frac{U_{max}-U_{min}}{(1-\delta)K\mathbb{E}\left[R_i\right]}.   \label{eq:delta_y} 
\end{equation}
Eq.(\ref{eq:delta_y}) shows that the amount of extra resources \textcolor{black}{needed} to obtain the maximum resolution is inversely proportional to the first moment of the per-block rate of the user. Consequently, to maximize the number of users who get the maximum resolution, we need to rank the users in decreasing order of their average per-block rates $\mathbb{E}[R_i]$. 
Following this reasoning, we have:
\begin{result}
The maximum number of users that can maintain the maximum resolution is
%\vspace{-6pt}
%\footnotesize
\begin{equation}
\text{max}\left\{N\middle|\sum_{i=1}^{N}\Delta Y_{(i)}\leq 1-\sum_{j=1}^{n}Y_{j,min}\right\},
\end{equation}
%\normalsize
where $\Delta Y_{(i)}$ is the extra frame ratio \textcolor{black}{the} resources \textcolor{black}{are} needed for the user with the $i$th highest average per-block rate to get the maximum playout rate.
\end{result}

\subsection{Two classes of users}
In Section~\ref{subsub:same} \textcolor{black}{we assumed that all the users %expected to 
have the same playout rate}, whereas in Section~\ref{subsec:ultimate} we were interested in maximizing the number of users with a maximum resolution by providing a minimum playout rate to everyone else. In a way, even in the second case there is only one class of users. Everyone is guaranteed the same minimum playout rate. The development of network slicing in 5G~\cite{p11} has allowed operators to split users into groups, with users of similar use cases or applications/services %of users 
within the same group (slice). Consequently, operators can split users into different classes (based on their QoE), as is currently the case with media-service providers like Netflix.   

We assume that there are two classes of users\footnote{Having more than two classes is straightforward with similar conclusions drawn. Therefore, to ease the presentation, we focus on two classes.}. Users \textcolor{black}{that are willing to pay more for the service have} a better QoE (higher playout rates) are called \emph{premium users}; all the rest (lower playout rates) are called \emph{regular users}. %The former will experience higher playout rates. %, and thus higher playout rates. 
Let $U_p$ be the playout rate for premium users, and $U_r$ the playout rate for regular users.  It holds that $U_p=k_pU_r, k_p\geq 1$. The outage in both cases is $\epsilon$. %dedicated to both classes of users. 
%Network slicing in 5G enables splitting these channels (blocks) between the premium and regular users. 
The outage probability for premium users is $\epsilon_p$, whereas for regular users it is $\epsilon_r=k_{\epsilon}\epsilon_p, k_{\epsilon}\geq 1$. %i.e., premium users suffer less rebuffering events. 
Similarly, the packet drop rate for premium users is $\delta_p$, and for regular users is $\delta_r=k_{\delta}\delta_p, k_{\delta}\geq 1$. So, premium users suffer less rebuffering events and lower information losses.

There are $K$ blocks in total. We assume that all the users (premium and regular) have the same buffer size $B$. Let $K_p$ be the number of blocks dedicated to premium users and $K_r$ the number of \textcolor{black}{blocks}  \textcolor{black}{for} regular users. There are two goals in front of the mobile operator. The first is to determine the maximum playout rates that can be guaranteed to both classes. The second goal is to determine the optimal assignment of blocks to both \textcolor{black}{groups}.    

Let $n_p$ be the number of premium users with per-block rates $R_{p,i}, i=1,\ldots,n_p$, and $n_r$ the number of regular users with per-channel rates $R_{r,j}, j=1,\ldots,n_r$. Since all the users within the class are to have the same playout rate, the data rate process is, as derived in Section~\ref{subsub:same}, $C_p(t)=K_pY_i(t)R_{p,i}(t)=\frac{K_p}{\sum_{i=1}^{n_p}\frac{1}{R_{p,i}(t)}}$ for premium users, and $C_r(t)=K_rY_j(t)R_{r,j}(t)=\frac{K_r}{\sum_{j=1}^{n_r}\frac{1}{R_{r,j}(t)}}$ for regular users. The procedure continues for both classes in line with Eqs.(\ref{eq:convolution_0})-(\ref{eq:arrival_process}). However, to get $K_p$ (or $K_r$) from the corresponding Eq.(\ref{eq:arrival_process}) is cumbersome and possible only numerically. Instead, we use the same approximation that lead to Eq.(\ref{eq:Y_min}) (for small $\epsilon$). In this case, this yields $\mathbb{E}\left[\frac{(1-\delta_p)K_p}{\sum_{i=1}^{n_p}\frac{1}{R_{p,i}}}\right]\approx U_p$ and $\mathbb{E}\left[\frac{(1-\delta_r)K_r}{\sum_{j=1}^{n_r}\frac{1}{R_{r,j}}}\right]\approx U_r$. The previous two equations lead to
%\vspace{-6pt}
%\footnotesize
%\begin{equation}
$K_p=\frac{U_p}{(1-\delta_p)\mathbb{E}\left[\frac{1}{\sum_{i=1}^{n_p}\frac{1}{R_{p,i}}}\right]}$ and 
%$$\label{eq:Kp}    
%$\end{equation}
%$$\normalsize
$K_r=\frac{U_r}{(1-\delta_r)\mathbb{E}\left[\frac{1}{\sum_{j=1}^{n_r}\frac{1}{R_{r,j}}}\right]}$. Dividing the last two equations, we have
%\vspace{-6pt}
%\footnotesize
\begin{equation}
%\frac{K_p}{K_r}=k_p\frac{E_1}{E_2},
\frac{K_p}{K_r}=k_p\cdot\frac{1-\delta_r}{1-\delta_p}\cdot\frac{E_1}{E_2}=c,   
\label{eq:Kp/Kr}
\end{equation}
%\normalsize
where $E_1=\mathbb{E}\left[\frac{1}{\sum_{j=1}^{n_r}\frac{1}{R_{r,j}}}\right]$, $E_2=\mathbb{E}\left[\frac{1}{\sum_{i=1}^{n_p}\frac{1}{R_{p,i}}}\right]$. %and $k_p=\frac{U_p}{U_r}$. 
Solving Eq.(\ref{eq:Kp/Kr}) together with $K=K_p+K_r$, and replacing the so obtained %values for 
$K_p$ and $K_r$ into the corresponding aforementioned \textcolor{black}{approximations} for $U_p$ and $U_r$, we get:
\begin{result}
The maximum playout rates that can be guaranteed to premium and regular users, \textcolor{black}{with outages $\epsilon_p$ and $\epsilon_r$}, respectively, and with drop rates of $\delta_p$ and $\delta_r$, respectively, are
%\vspace{-6pt}
%\footnotesize
\begin{equation}
%U_{p,max}=\frac{k_1K\mathbb{E}\left[\frac{1}{\sum_{i=1}^{n_p}\frac{1}{R_{p,i}}}\right]\mathbb{E}\left[\frac{1}{\sum_{j=1}^{n_r}\frac{1}{R_{r,j}}}\right]}{k_1\mathbb{E}\left[\frac{1}{\sum_{j=1}^{n_r}\frac{1}{R_{r,j}}}\right]+\mathbb{E}\left[\frac{1}{\sum_{i=1}^{n_p}\frac{1}{R_{p,i}}}\right]}, and
%U_{p,max}=\frac{k_pKE_1E_2}{k_pE_1+E_2}, and
U_{p}=(1-\delta_p)\cdot\frac{cK}{1+c}E_2,
\label{eq:Up}
\end{equation}
\begin{equation}
%U_{r,max}=\frac{K\mathbb{E}\left[\frac{1}{\sum_{i=1}^{n_p}\frac{1}{R_{p,i}}}\right]\mathbb{E}\left[\frac{1}{\sum_{j=1}^{n_r}\frac{1}{R_{r,j}}}\right]}{k_1\mathbb{E}\left[\frac{1}{\sum_{j=1}^{n_r}\frac{1}{R_{r,j}}}\right]+\mathbb{E}\left[\frac{1}{\sum_{i=1}^{n_p}\frac{1}{R_{p,i}}}\right]}.
%U_{r,max}=\frac{KE_1E_2}{k_pE_1+E_2}.
U_{r}=(1-\delta_r)\cdot\frac{K}{1+c}E_1.
\label{eq:Ur}    
\end{equation}
%\normalsize
\end{result}
\emph{Note}: %\footnotesize 
$\mathbb{E}\left[\frac{1}{\sum_{i=1}^{n}\frac{1}{R_i}}\right]=\sum_{k_1=1}^{m}\cdots\sum_{k_n=1}^{m}\frac{1}{\sum_{i=1}^{n}\frac{1}{r_{k_i}}}\prod_{i=1}^{n}p_{R_i}(r_{k_i})$. %\normalsize. 

\subsection{Implementation}
\textcolor{black}{The advantage of our approach is that it can be implemented on top of any of the current 5G architectures~\cite{trivisonno2015towards},~\cite{gupta2015survey}. Essentially, there is only one extra step on the base station side for the resource allocation process (that of calculating the frame ratio for every user), depending on what the optimization objective is. 
Namely, every mobile user sends to the base station the Channel Quality Indicator (CQI), i.e., its per-block rate in the frame. Based on this information, when the goal is to provide the same experience to everyone,  
the base station calculates using Eq.(\ref{eq:Yi_equal}) the amount of resources for every user and allocates them. Similar steps are performed for the other problems.}

\section{Benchmark model and QoE metrics}
\label{sec:qoe}
\textcolor{black}{The approach we present in this work relies on the requirement that every user is to be guaranteed a fixed playout rate throughout the streaming process. In order to show the usefulness of this approach to user experience, we compare the performance with benchmarks.} 

It is well known that that DASH supports adaptive bitrate streaming (ABR)~\cite{Xu14_buffer_thresholds}. %Also, because its commercial character 
Because the DASH algorithm is not public,
it is not fully known how the playout rate changes. However, it is known~\cite{Xu14_buffer_thresholds} that the playout rate changes according to the state of the buffer. To capture this behavior, we use an approach very similar  to~\cite{Xu14_buffer_thresholds} to model the change in the playout rate depending on the state of the buffer. There are two characteristic levels of the state of the buffer: $B_{min}$ and $B_{max}$. While the state of the buffer is in the range $[B_{min},B_{max}]$ the playout rate will remain unchanged. When the state of the buffer drops below $B_{min}$, the playout rate decreases by $\theta$\%, deteriorating the video resolution. If the state of the buffer goes above $B_{max}$ the playout rate increases by $\theta$\%. In Section~\ref{sec:evaluation}, we run simulations for different values of $B_{min}$, $B_{max}$, and $\theta$. In order to compare the performance of such a system against our approach, we need to define the QoE metrics according to which the comparison will be conducted.  

There is no general consensus regarding the most suitable metric for the QoE of mobile users that stream live videos. In general, QoE metrics can be classified as the ``simple'', such as: average playout rate, the number of rebuffering events, the drop rate, etc., or as ``composite'', in which the QoE is a function of the simple metrics. In this paper, we consider one of these metrics, which is defined as %three from the latter group. The first is from~\cite{NOVA}, and is defined as
\begin{equation}
QoE_i=\mathbb{E}[U_i]-\eta_i\text{Var}(U_i),
\label{eq:nova}    
\end{equation}
where $U_i$ is the playout rate over time, and $\eta_i>0$ scales penalty for temporal variability in quality~\cite{NOVA}.  

%\textcolor{red}{The second QoE metric is from~\cite{Feng20}, and is defined as
%\begin{equation}
%QoE_i=-\delta\sum_{n=1}^{N}|R_{n+1}-R_n|+\sum_{n=1}^{N}\sum_{f=1}(\alpha R_{n,f}-\beta T_{n,f}-\gamma L_{n,f}),
%\label{eq:vabis}    
%\end{equation}
%where $R_n$ denotes the bitrate of the $n$th decision point, $R_{n,f}$ is the bitrate of the $f$th frame between the $n$th decision points and the $(n+1)$th decision points, $T_{n,f}$ represents the rebuffering time that results from downloading the $f$th frame at playout rate $R_{n,f}$. The parameter $L_{n,f}$ denotes the latency that results from downloading the $f$th frame at $R_{n,f}$. The coefficient $\alpha$ represents the video length of each frame, whereas $\beta$, $\gamma$, and $\delta$ denote the penalty weight of the rebuffering, latency and bitrate switching, respectively.}    

%\textcolor{black}{The third QoE metric that we consider in this work is from~\cite{Guohong_icdcs}, but due to space limitations we refer the interested reader to our tech report~\cite{XX} for more details on that QoE metric.}   

%and is defined as
%\begin{equation}
%QoE_i=\max (1,\min(5,1+4\frac{c_1 b}{c_2+b}))-f_rI_r-f_bI_b-c_3-c_4\exp(c_5bv),
%\label{eq:guohong_icdcs}    
%\end{equation}
%where $b$ is the bitrate, $c_1$ and $c_2$ are the model parameters that are determined by the subjective quality assessment experiments, $I_r=0.742$, $f_r$ is the rebuffering frequency, $v$ is the vibration level, $c_3$, $c_4$, and $c_5$ are the model parameters.  

\section{Performance evaluation}
\label{sec:evaluation}
In this section, we first describe the simulation setup. Then, we provide validations of (\ref{eq:U_c_finite}) and of approximation result (\ref{eq:Y_min}) using trace data. We then look at how our theoretical result predicts the performance for packet sizes that are not deterministic.    
This is followed by results related to the three problems.  After that, we compare the performance of our approach with that of an adaptive bitrate (ABR) system for \textcolor{black}{four} QoE metrics. \textcolor{black}{Finally, we look at the impact of buffer sizes on the packet drop rate.} 

%Finally, we show the impact of playout rate on the number of users that can be admitted. %The simulations are run in MATLAB R2018b.   
%This is followed by results on the amount of resources left unused after providing consistency. 
%Then, we present outcomes on throughput optimization and fairness-related results. Finally, we show some results related to two classes of users. \textcolor{black}{All the simulations are conducted in Matlab.} 

%\subsection{Simulation setup}
%\textcolor{black}{We present first the synthetic-driven setup. This is followed by the setup in which the input parameters are obtained from a real-life trace.} 

\subsection{Simulation setup}
As input parameters, we have used data from a trace of the signal quality of mobile users. These traces can be found in~\cite{trace}, and their detailed description is provided in~\cite{trace_paper}.\footnote{\textcolor{black}{Due to the lack of publicly-available 5G traces, we used the signal quality from 4G measurements. Nevertheless, the received signal powers encountered in 4G and 5G are similar~\cite{3GPP_5G_NR}, with all other parameters being 5G-related.}} The measurements were conducted in several cities  \textcolor{black}{across} Europe and North America. Among the parameters we are interested in from the trace are the Received Signal Strength Indicator (RSSI) and users' positions, where the latter \textcolor{black}{are} expressed in terms of their longitude and latitude. We picked 8 users in Amsterdam, which were chosen based on their positions to be close enough so that they can be served by the same gNodeB. %Then, for every user and the corresponding RSSI values, 
Then, 
we mapped RSSI values of every user over time to the corresponding SINR values taken from~\cite{baicells}. %\textcolor{black}{choosing the  values that were closer to our setup}. 
\textcolor{black}{We chose $m=15$, which means that all SINR values were translated into 15 discrete per-block rates (second row of Table~\ref{Table-R_2}), according to the threshold values $\gamma$\cite{SINR_threshold}, shown in the first row of Table~\ref{Table-R_2}. E.g., if in a frame a user's SINR is $7$ dB, its per-block rate in that frame is $712$ kbps.} 
%The 
%SINR values for every user were then translated to the corresponding per-block rates (second row of Table~\ref{Table-R_2}).  
Further, based on the frequency of occurrence of a per-block rate for every user, we obtained per-block rate probabilities in Table~\ref{Table-R_2}. \textcolor{black}{From the trace, we have observed a strong correlation between the received signals of a user in contiguous frames.}   

The frame duration is $10$\:ms. \textcolor{black}{The subcarrier spacing is $30$\:KHz, with $12$ subcarriers per block, making the block width $360$\:KHz.} %\footnote{\textcolor{blue}{There are 3 subcarrier spacing options in sub-6 GHz 5G: $15$, $30$, and $60$\:KHz, as opposed to 4G where only the option of $15$\:KHz exists, with the number of subcarriers the same as in 5G. In our case, although spectral efficiencies are the same in 4G and 5G~\cite{3GPP_5G_NR}, \textcolor{black}{ because of $2\times$ higher subcarrier spacing,} the per-block rates in Table~\ref{Table-R_2} are $2\times$ higher than in LTE.}}} %with a total bandwidth of 50 MHz. Hence, 
%There are $K=\lfloor{\frac{50}{0.36}\rfloor}$ blocks
\textcolor{black}{The total number of PRBs is $K=275$~\cite{3GPP_5G_NR}}. The size of a packet is  $5$\:kbits.\footnote{\textcolor{black}{We tried other values for  packet sizes with similar conclusions drawn.}} \textcolor{black}{The duration of the event is $2.5$ hours.} \textcolor{black}{Unless stated otherwise, the buffer size is $3$\:MB.} 

\textcolor{black}{All the simulations are conducted in MATLAB R2018b and we take the average of the metrics of interest over 1000 runs.}
\begin{table}[t]
\caption{Per-block rates and the corresponding probabilities for every user from the sampled Amsterdam trace}
\label{Table-R_2}
\scriptsize
%\centering
%\footnotesize
\begin{tabular}{|c|c|c|c|c|c|c|c|c|c|c|c|c|c|c|c|}
\hline
\hspace{-6pt}SINR(dB)\hspace{-6pt} &\hspace{-6pt} -9.5 \hspace{-6pt}&\hspace{-6pt} -6.7 &\hspace{-6pt} -4.1 &\hspace{-6pt} -1.8 & 0.4 & 2.4 & 4.5 & 6.4 & 8.5 & 10.3 & 12.2 & 14.1 & 15.8 & 17.8 & 19.8\\
\hline
\hspace{-6pt}R(kbps) &\hspace{-6pt} 48 \hspace{-6pt}&\hspace{-6pt} 73.6 &\hspace{-6pt} 121.8 &\hspace{-6pt} 192.2 & 282 & 378 & 474.2 & 712 & 772.2 & 874.8 & 1063.8 & 1249.6 & 1448.4 & 1640.6 & 1778.4\\
\hline
\hspace{-10pt}$p_{1,k}$ &\hspace{-6pt} 0 \hspace{-6pt}&\hspace{-6pt} 0.1 &\hspace{-6pt} 0.72 &\hspace{-6pt} 0.04 & 0.05 & 0.09 & 0 & 0 & 0 & 0 & 0 & 0 & 0 & 0 & 0\\
\hline   
\hspace{-10pt}$p_{2,k}$ &\hspace{-6pt} 0 \hspace{-6pt}&\hspace{-6pt} 0 &\hspace{-6pt} 0.2 &\hspace{-6pt} 0.7 & 0.1 & 0 & 0 & 0 & 0 & 0 & 0 & 0 & 0 & 0 & 0\\
\hline
\hspace{-10pt}$p_{3,k}$ &\hspace{-6pt} 0 \hspace{-6pt}&\hspace{-6pt} 0 &\hspace{-6pt} 0 &\hspace{-6pt} 0 & 0.02 & 0.12 & 0.51 & 0.32 & 0.01 & 0.01 & 0.01 & 0 & 0 & 0 & 0\\
\hline
\hspace{-10pt}$p_{4,k}$ &\hspace{-6pt} 0 \hspace{-6pt}&\hspace{-6pt} 0 &\hspace{-6pt} 0 &\hspace{-6pt} 0 & 0 & 0.01 & 0.98 & 0.01 & 0 & 0 & 0 & 0 & 0 & 0 & 0\\
\hline                                                         
\hspace{-10pt}$p_{5,k}$ &\hspace{-6pt} 0.22 \hspace{-6pt}&\hspace{-6pt} 0.04 &\hspace{-6pt} 0.07 &\hspace{-6pt} 0.04 &  0.04 & 0.06 & 0.17 & 0.15 & 0.01 & 0.01 & 0.06 & 0.06 & 0 & 0.03 & 0.04\\
\hline 
\hspace{-10pt}$p_{6,k}$ &\hspace{-6pt} 0.17 \hspace{-6pt}&\hspace{-6pt} 0.11 &\hspace{-6pt} 0.1 &\hspace{-6pt} 0.07 & 0.05 & 0.1 & 0.17 & 0.11 & 0.02 & 0.04 & 0 & 0.03 & 0 & 0.02 & 0.01\\
\hline                                                         
\hspace{-10pt}$p_{7,k}$ &\hspace{-6pt} 0.05 \hspace{-6pt}&\hspace{-6pt} 0.03 &\hspace{-6pt} 0.06 &\hspace{-6pt} 0.07 &  0.09 & 0.17 & 0.33 & 0.08 & 0.01 & 0.01 & 0.01 & 0.03 & 0.01 & 0.03 & 0.02\\
\hline
\hspace{-10pt}$p_{8,k}$ &\hspace{-6pt} 0 \hspace{-6pt}&\hspace{-6pt} 0 &\hspace{-6pt} 0 &\hspace{-6pt} 0.02 &  0.01 & 0.03 & 0.06 & 0.08 & 0.01 & 0.02 & 0.01 & 0.03 & 0 & 0.05 & 0.68\\
\hline
\end{tabular}
\normalsize
%\vspace{-9pt}
\end{table}

\subsection{\textcolor{black}{Validation results}}
%We first validate Eq.(\ref{eq:U_c}), and then Eq.(\ref{eq:Y_min}).
\subsubsection{Validating the playout rate result} We validate the result for the maximum \textcolor{black}{achievable} playout rate for different values of the outage and two users, user 1 and user 8. The resources are split equally among the 8 users. The data rate for these users in frame $t$ is $\frac{K}{8}R_1(t)$ and $\frac{K}{8}R_{8}(t)$, respectively.  %We show the results for user 1 and user 8. 
%In the first case, the data rate is random and chosen uniformly in the range $[5,10]$ Mbps, whereas in the second it uniform in $[2,4]$ Mbps. 
Table~\ref{table:validation_1} depicts the highest achievable playout rates that can be guaranteed to the two users for different maximum allowed $\epsilon$, \textcolor{black}{when $\delta_0=0.03$}. \textcolor{black}{There are several interesting outcomes of these results. \textcolor{black}{Firstly and most importantly}, it can be observed that our theoretical result closely matches the simulation results, despite the fact that in our approach we assume that per-block rates \textcolor{black}{of the same user} in contiguous frames are independent, whereas in the trace there is a correlation. Secondly, when increasing the outage by small amounts there is no considerable gain in playout rates. %The large buffer is the reason for providing high playout rates even when the outage is low. 
Thirdly, user 8 can afford, due to better channel conditions, much higher playout rates.}   
\begin{table}
\centering
\caption{Playout rate validation for $\delta_0=0.03$}\label{table:validation_1}
%\vspace{-6pt}
\begin{tabular}{|l|l|l|l|l|l|} \hline
 Outage $(\epsilon$) & 0.01 & 0.03 & 0.05 & 0.08 & 0.1 \\\hline
 User 1-theory (Mbps) & 5 & 5.06 & 5.12 & 5.23 & 5.29  \\\hline 
 User 1-sims (Mbps) & 5.02 & 5.09 & 5.15 & 5.21 & 5.35 \\\hline
 User 8-theory (Mbps) & 50.37 & 50.89 & 51.48 & 52.27 & 53.26\\\hline
 User 8-sims  (Mbps) & 50.09 & 50.69 & 51.22 & 52.46 & 53.43\\\hline
 \end{tabular}
%\vspace{-10pt}
\end{table}

\textcolor{black}{Next, Table~\ref{table:validation_1_0} shows the theoretical (Eq.(\ref{eq:U_c_finite})) and simulation results for different maximum allowed values of $\delta_0$, when $\epsilon=0.05$. We show the results for user 3 and user 6. Similarly to the previous scenario, as can be observed from Table~\ref{table:validation_1},  there is a close match between the theory and actual results. The discrepancy never exceeds $2-3$\%. Relaxing the requirement on the maximum allowed drop rate leads to lower required playout rates to support the same outage because increasing the playout rate decreases the number of dropped packets.} %The drop rate decreases with the increase of playout rate.} 
\begin{table}
\centering
\caption{Playout rate validation for $\epsilon=0.05$}\label{table:validation_1_0}
%\vspace{-6pt}
\begin{tabular}{|l|l|l|l|l|l|} \hline
Max. drop rate $(\delta_0$) & 0.01 & 0.03 & 0.05 & 0.08 & 0.1 \\\hline
 User 3-theory (Mbps) & 17.81 & 17.35 & 17.02 & 16.46 & 16.11  \\\hline 
 User 3-sims (Mbps) & 17.48 & 17.22 & 16.98 & 16.5 & 15.98 \\\hline
 User 6-theory (Mbps) & 13 & 12.75 & 12.25 & 11.75 & 11.5\\\hline
 User 6-sims  (Mbps) & 12.98 & 12.77 & 12.14 & 11.78 & 11.44\\\hline
 \end{tabular}
%\vspace{-10pt}
\end{table}

\begin{table}
\centering
\caption{Frame ratio validation for $\epsilon=0.01$ and $\delta_0=0.04$}\label{table:validation_2}
%\vspace{-6pt}
\begin{tabular}{|l|l|l|l|l|} \hline
 Minimum playout (Mbps)  & 4 & 8 & 20 & 40 \\\hline
  User 2-theory ($Y_{2,min}$) & $0.082$ & $0.163$ & $0.408$ & $0.817$\\\hline
 User 2-sims  ($Y_{2,min}$) & $0.079$ & $0.158$ & $0.39$ & $0.791$ \\\hline
 User 7-theory ($Y_{7,min}$) & $0.032$ & $0.063$ & $0.158$ & $0.316$\\\hline 
 User 7-sims ($Y_{7,min}$) & $0.031$ & $0.057$ & $0.156$ & $0.309$  \\\hline
 \end{tabular}
%\vspace{-10pt}
\end{table}

\subsubsection{Validating the minimum \textcolor{black}{guaranteed} playout rate result} We validate the accuracy of our approximation result (Eq.(\ref{eq:Y_min})). To this end, the minimum playout \textcolor{black}{rates are} $U_{min}=\{4,8,20,40\}$\:Mbps, and there are two users whose results we show, users 2 and 7. For both users, $\epsilon=0.01$ and $\delta_0=0.04$.  Table~\ref{table:validation_2} shows the results of frame ratio $Y$ needed to provide $U_{min}$. The actual results (marked as ``sims'') are obtained numerically by the \emph{trial-and-error} method until the exact value of $Y_{i,min}$ that provides $U_{min}$ is obtained. As can be observed, our approximation (Eq.(\ref{eq:Y_min})) closely matches the actual result. %for $\epsilon=0.01$ \textcolor{black}{and $\delta=0.05$}. 
For higher values of $\epsilon$ (results not shown due to space limitations) the discrepancy is on the order of $10$\%. This proves the usefulness of our approach in practice. Due to better channel conditions, user 7 needs fewer resources to maintain the same playout rate.   

\subsubsection{Variable packet sizes}
Having validated the results for the playout rate and the frame ratio, we proceed with validating our theoretical result for the playout rate %(with fixed packet sizes of $5$\:kb)
(which holds for fixed packet sizes)
for the case when the packet size is not constant. To this end, we assume that packet sizes are uniform in the range $[4,6]$\:kb. The other parameters are the same as in the previous scenarios. User 4 and user 5 are our users of interest in this scenario. Fig.~\ref{fig:uniform} depicts the maximum playout rate for the video vs. drop rate for the actual result (uniform packet sizes) and that predicted by our theory (which pertains to deterministic packet sizes with the same value as the average of the uniform packets, i.e., $5$\:kb). The results correspond to $\epsilon=0.05$. As can be seen from Fig.~\ref{fig:uniform}, our theory fares pretty well for non-deterministic packet sizes as well, with the level of discrepancy not exceeding $5$\%. This shows the practical usefulness of our model. User 5 has better channel conditions, and as a result of that, a higher playout rate.

\subsection{Optimization results}
%Having validated the results, we proceed with the corresponding optimal results. 
We first illustrate a scenario related to providing the same playout rate to everyone. This is followed by a scenario related to maximizing the number of users with a given resolution and a scenario with two classes of users.

\subsubsection{Same experience to everyone}
\begin{figure}
 \begin{minipage}{0.48\linewidth}
    \centering
  \includegraphics[width=\linewidth]{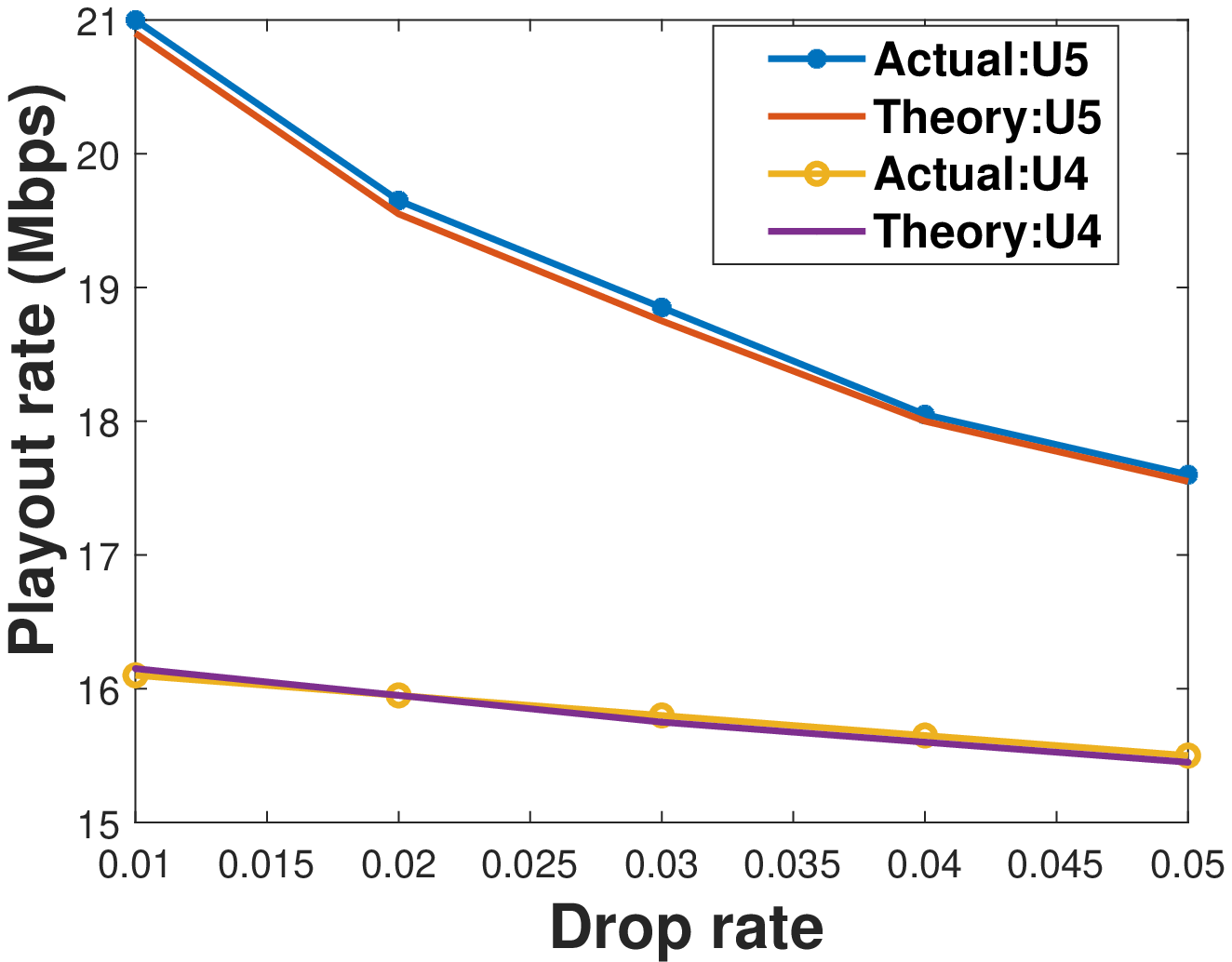}
   \caption{$U_p$ vs. drop rate for variable packet sizes.}
    \label{fig:uniform}
   \end{minipage}
\begin{minipage}{0.48\linewidth}
\centering
\includegraphics[width=\textwidth]{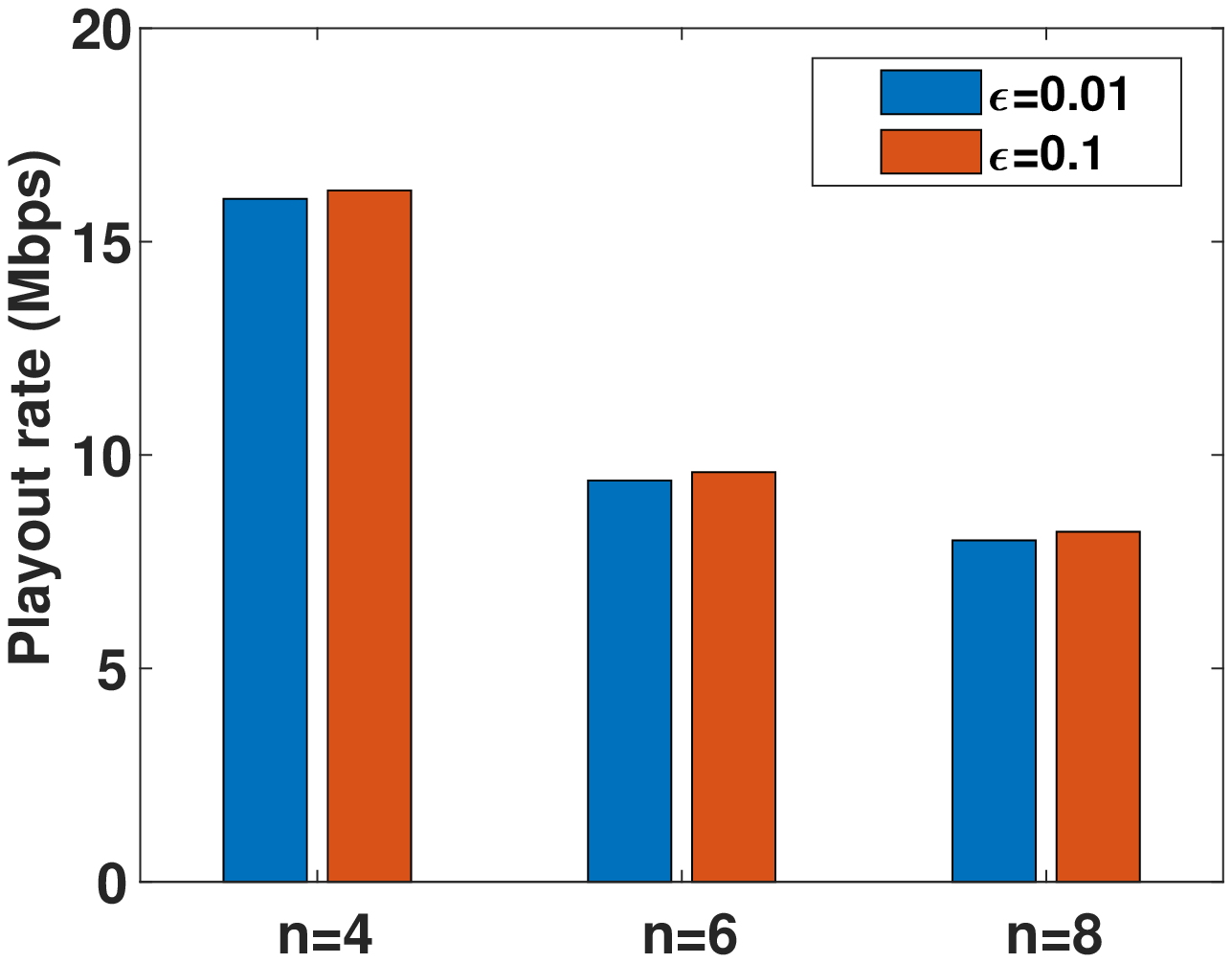}
\caption{The impact of number of users on playout rate.}
\label{bar_playout}
\end{minipage}
\end{figure}

%Having validated the results, we proceed with the corresponding optimal results. 
The first goal is to determine the maximum achievable playout rate for different number of users and different values of $\epsilon$. %\textcolor{black}{and $\delta$}. 
There are three scenarios: with users 1-4, users 1-6, and users 1-8. Fig.~\ref{bar_playout} shows the maximum achievable playout rate vs. $\epsilon$, \textcolor{black}{when $\delta_0=0.01$.} As expected, the more users there are, the lower the playout rate is. Again, allowing higher outages provides small gains in playout rates - \textcolor{black}{order of $5$\% at most}.  
%\begin{minipage}{0.24\linewidth}
%\centering
%\includegraphics[width=\textwidth]{buffer_2_4.eps}
%\caption{The evolution of the state of the buffer.}
%\label{buffer}
%\end{minipage}
%\begin{minipage}{0.24\linewidth}
%\centering
%\includegraphics[width=\textwidth]{outage_vs_buffer_2.eps}
%\caption{The impact of reducing latency on outage.}
%\label{outage_vs_latency}
%\end{minipage}
\begin{figure}
\begin{minipage}{0.48\linewidth}
\centering
\includegraphics[width=\textwidth]{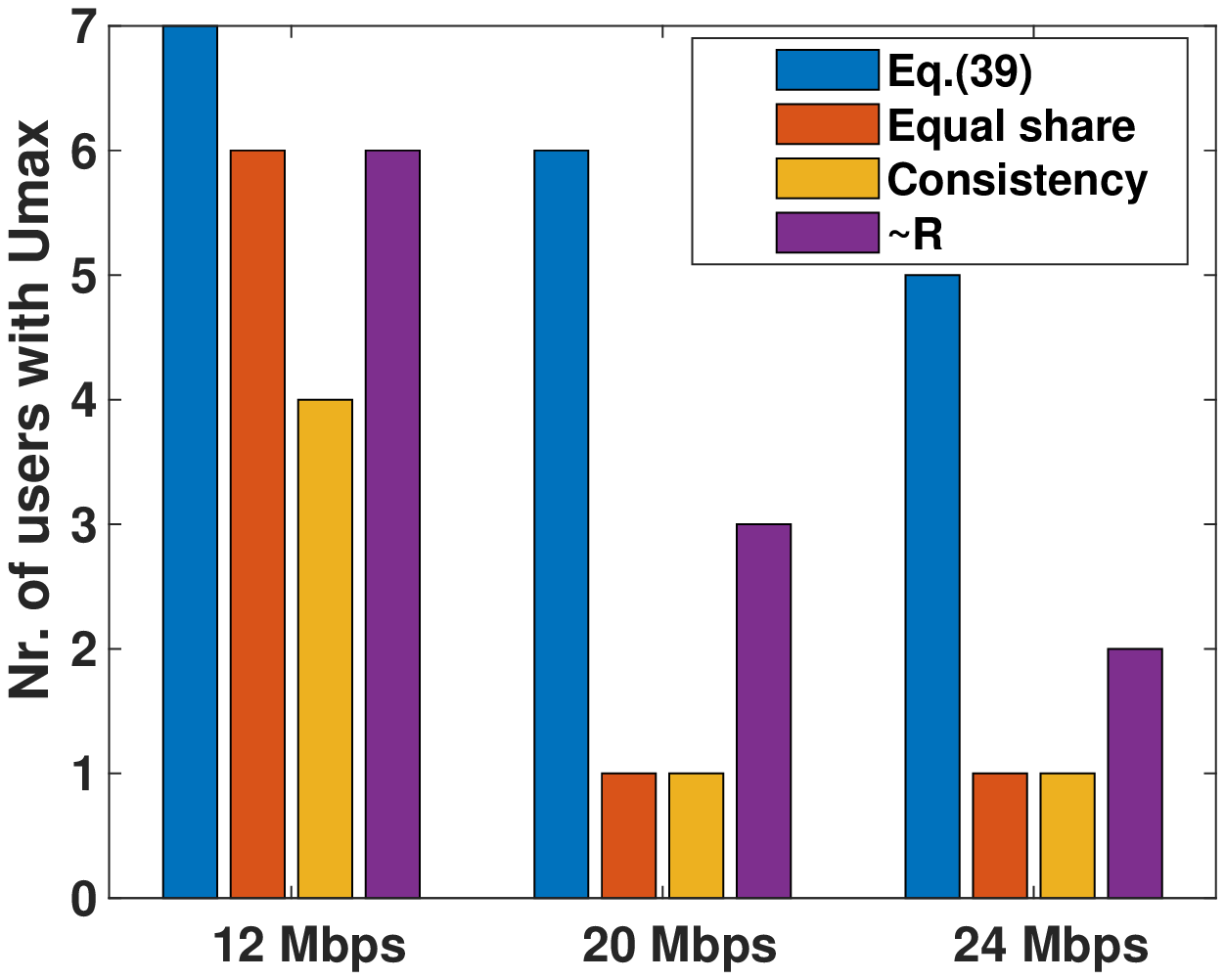}
\caption{Number of users with maximum playout rate.}
\label{bar_maximum_resolution}
\end{minipage}
 \begin{minipage}{0.48\linewidth}
    \centering
  \includegraphics[width=\linewidth]{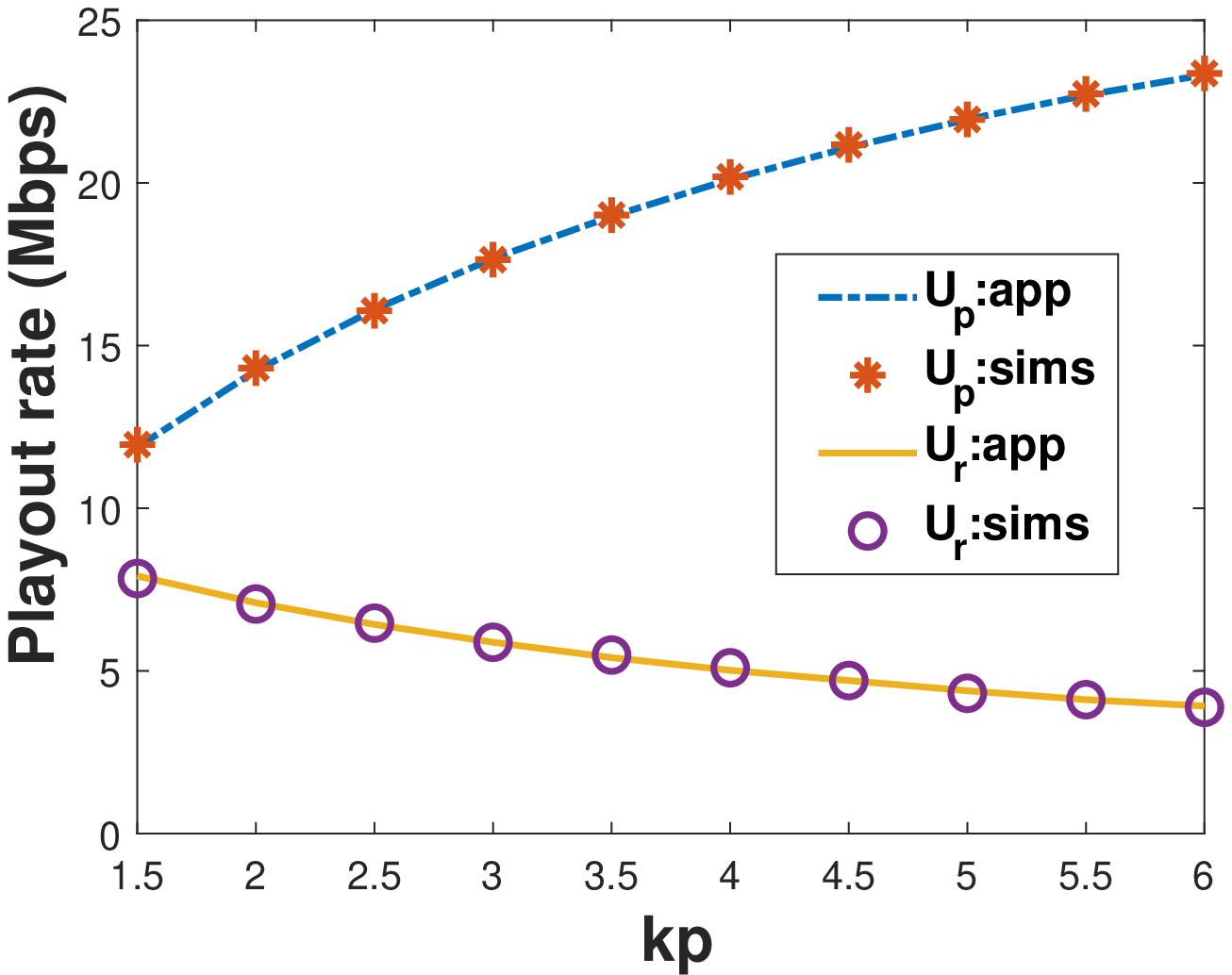}
   \caption{The playout rate for two classes of users.}
    \label{two_classes_playout}
   \end{minipage}
%\vspace{-10pt}
\end{figure}
\begin{figure}
\begin{minipage}{0.48\linewidth}
    \centering
  \includegraphics[width=\linewidth]{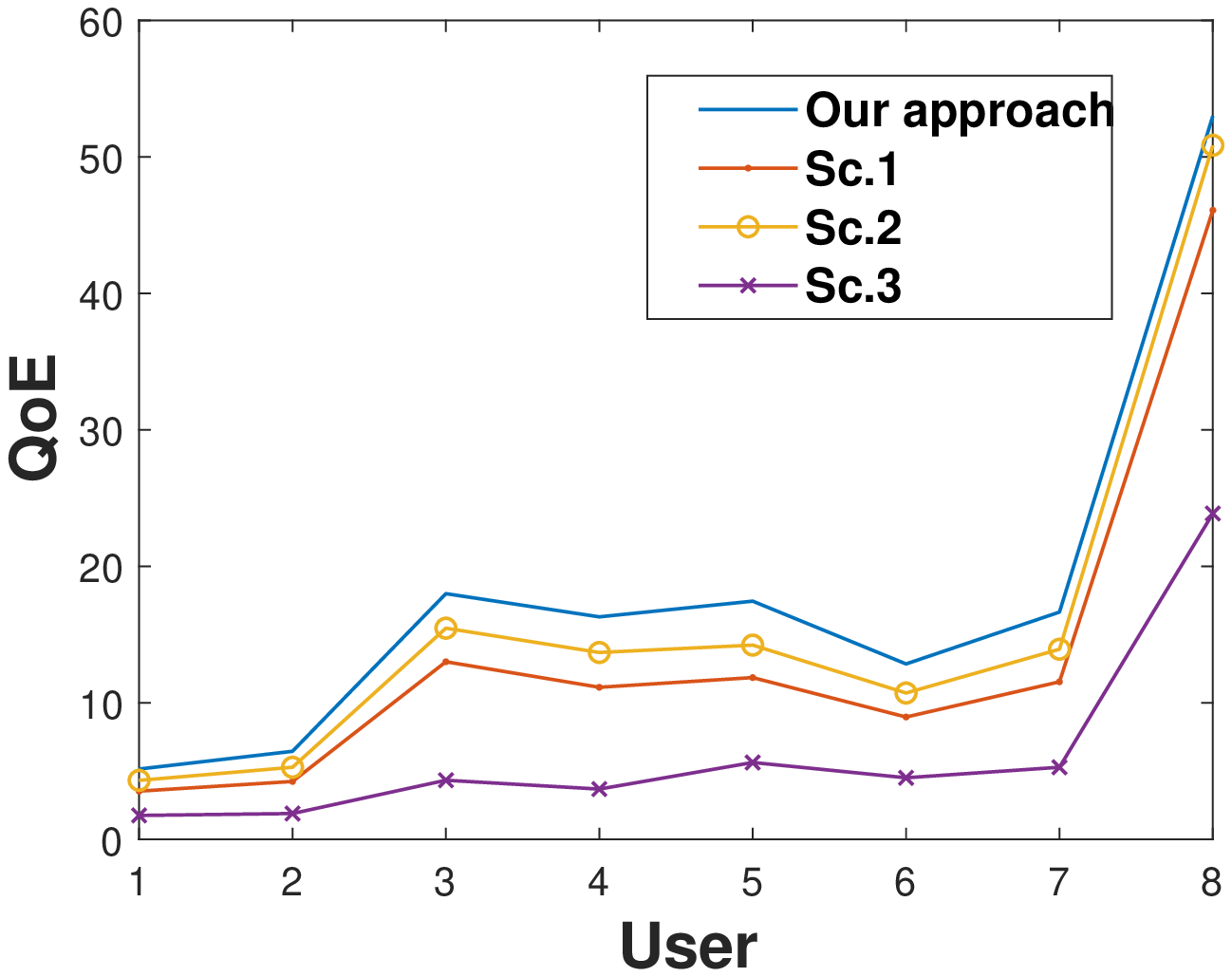}
   \caption{QoE for our approach and ABR.}
    \label{fig:qoe}
   \end{minipage}
\begin{minipage}{0.48\linewidth}
\centering
\includegraphics[width=\linewidth]{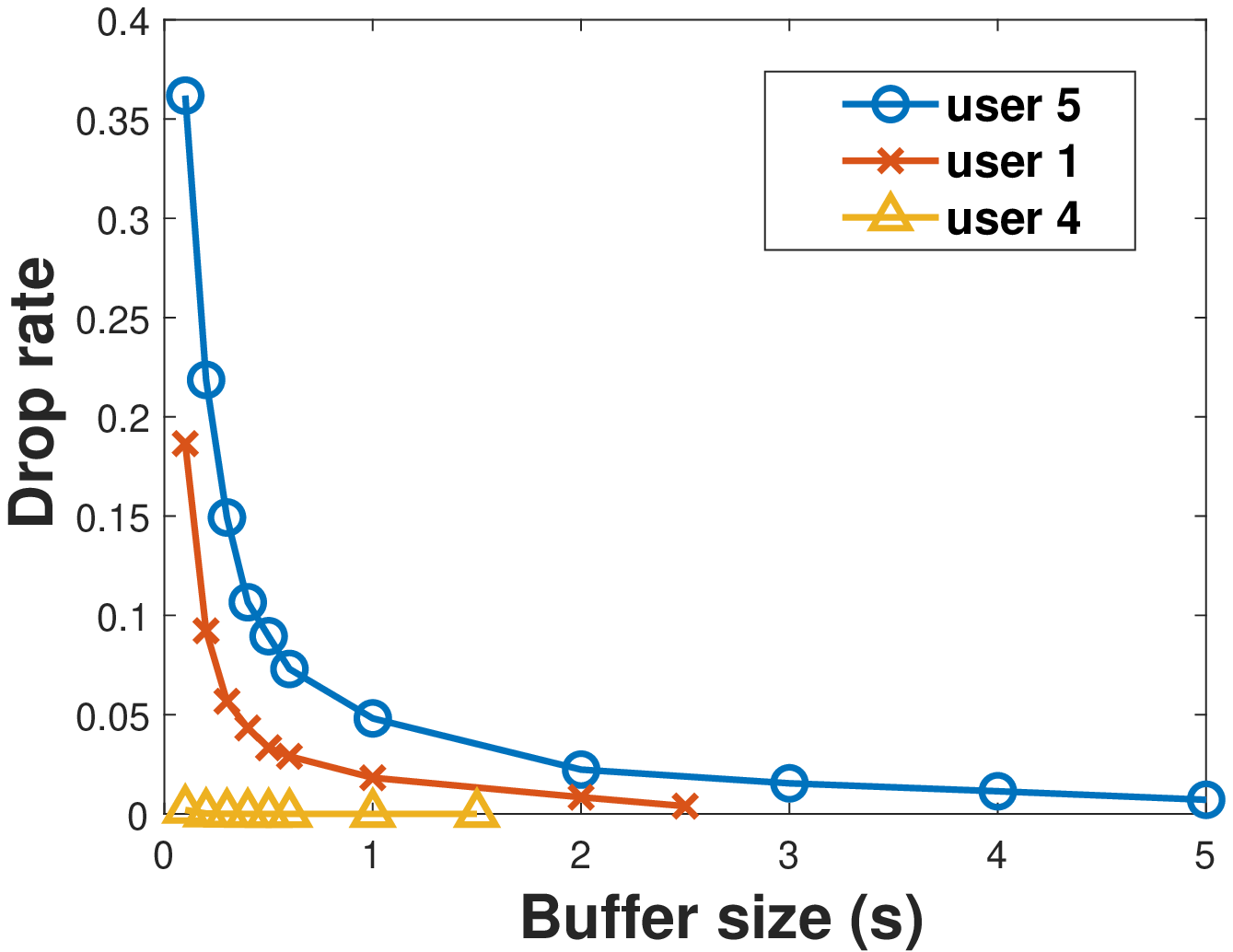}
\caption{Drop rate as a function of buffer size.}
\label{fig:buffer_seconds}
\end{minipage}
%\begin{minipage}{0.48\linewidth}
 %   \centering
 % \includegraphics[width=\linewidth]{latency_buffer_finite.eps}
  % \caption{The latency for user 5.}
  %  \label{fig:buffer_latency}
  % \end{minipage}
%\begin{minipage}{0.48\linewidth}
%\centering
%\includegraphics[width=\textwidth]{admission_2.eps}
%\caption{Maximum \textcolor{black}{number of users} that can have a \textcolor{black}{given} playout rate.}
%\label{admission}
%\end{minipage}
%\vspace{-10pt}
\end{figure}

\subsubsection{Maximum playout rate} %In this scenario, we replicate the 8 users once, which means that now there are 16 users. 
We compare next the number of users that achieve a given (maximum) resolution within the pool of the considered 8 users % which in this case is $5$ Mbps
following our policy of Section~\ref{subsec:ultimate}
 with the results obtained from three other policies: a) where there is an equal share of the resources to everyone, %b) MaxWeight algorithm~\cite{}, which in a frame allocates all the resources to the user with highest $QR$ (queue $\times$ per-block rate), 
b) where the same constant data rate is provided to everyone at all times and then the unused resources are reallocated equally to the same users according to~\cite{Fidan_TMC}, and c) the resources are allocated proportionally to user's per-block rate, i.e.,  $Y_i(t)=\frac{R_i(t)}{\sum_{j=1}^{n}R_j(t)}$. We do this for three scenarios: $U_{max}=\{12,20,24\}$\:Mbps.\footnote{\textcolor{black}{This value is decided by the operator, and need not be the rate yielding the ultimate video experience.}} The minimum playout rate that must be guaranteed to every user in the cell is $U_{min}=2$\:Mbps. 

Fig.~\ref{bar_maximum_resolution} illustrates the results for %different values of $U_{max}$ and 
$\epsilon=0.01$ \textcolor{black}{and $\delta_0=0.03$}. As can be observed from Fig.~\ref{bar_maximum_resolution}, our policy provides the best result in all scenarios, outperforming the next best policy by at least $15$\%. E.g., following our policy, $7$ out of the $8$ users can be guaranteed a playout rate of $12$\:Mbps for $99$\% of the time, with $3$\% of the packets dropped.  

\subsubsection{Two classes of users} %As shown in Section~\ref{sec:optimization}, the policy that provides proportional fairness is the equal-share policy. Hence, we consider how our policy fares in terms of the objective (\ref{eq:equi_objective}) against the MaxWeight policy and consistent policy, introduced in the previous scenario. The parameters remain unchanged. Fig.~\ref{figure_5} illustrates the results for different $\epsilon$. In all the cases, our approach provides the highest objective.   
\textcolor{black}{We proceed with the case of two classes of users to see how our approximations fare against actual results. In this scenario, we pick users 1-5 to be regular, whereas users 6-8 are premium. The goal is to guarantee all the regular users the same playout rate with the same outage and drop rate. The same holds for premium users, whose rate is $k_p\times$ higher than of regular users. The outage for both classes is $\epsilon_p=\epsilon_r=0.1$, i.e., $k_{\epsilon}=1$, \textcolor{black}{whereas the drop rate is $\delta_p=\delta_r=0.01$ ($k_{\delta}=1$).}  Fig.~\ref{two_classes_playout} illustrates the approximation results for $U_p$ (Eq.(\ref{eq:Up})) and $U_r$ (Eq.(\ref{eq:Ur})) against actual (simulated) results for different $k_p$. The first thing to observe is the close match, with the discrepancy not exceeding $3$\%. 
\textcolor{black}{Secondly, playout rates of premium (regular) users
increase (decrease) slower than linearly with $k_p$. This conclusion propagates across other combinations of $k_{\epsilon}$ and $k_{\delta}$ as well.}} 
%Secondly, the playout rates of premium users in all the corresponding cases are $k_p\times$ higher. %Thirdly, the increase in premium playout rates is much more emphasized than the decrease in regular users playout rates. Namely, from     %Premium users will have a rate $k_1$ times higher than regular users.  

\subsection{QoE comparisons with ABR streaming}
\textcolor{black}{Next, we compare the performance of our approach (providing a constant playout rate throughout the entire streaming process) with the adaptive bitrate (ABR) streaming approach described in Section~\ref{sec:qoe}. For the latter, we consider three scenarios in terms of $B_{min}$, $B_{max}$, and $\theta$:  %we mimic to the closest possible extent the system from~\cite{Feng20}, with $B_{min}=0.25B$, $B_{max}=0.75B$, and $\theta=0.1$ (see Section~5). 
in the first scenario, $B_{min}=0.25B$, $B_{max}=0.75B$, $\theta=10$\%; in the second, $B_{min}=0.3B$, $B_{max}=0.8B$, $\theta=5$\%; and in the third scenario, $B_{min}=0.35B$, $B_{max}=0.85B$, $\theta=20$\%. In all the scenarios the resources are shared equally among all the users, i.e., every user will get $1/8$ of the resources.} Note that $B=\frac{3\:MB}{5\:kb}=\frac{24\:Mb}{5\:kb}=4800$ packets.

\textcolor{black}{We perform the comparison across three QoE metrics. The first QoE metric is  
%To that end, we use the QoE metric 
Eq.(\ref{eq:nova}), with $\eta=0.05$~\cite{NOVA}. The QoE for a user in our system is the playout rate itself (which is fixed). It has been obtained for $\delta=0.03$ and $\epsilon=0.01$ across all eight users.  Fig.~\ref{fig:qoe} shows the values of the QoE metric for all the eight users from the Amsterdam trace both for our approach and the ABR system (three scenarios). As can be observed from Fig.~\ref{fig:qoe}, the QoE with our system (constant playout rate) is higher than the QoE with the three scenarios pertaining to the ABR system over all users. %The same conclusion propagates over all values of $B_{min}$, $B_{max}$, and $\theta$. 
Hence, the advantages our approach provides.}       

\textcolor{black}{The second QoE metric is the packet drop rate $\delta$. Table~\ref{table:abr} shows the value of $\delta$ across all users for the three aforementioned scenarios. As already mentioned in the previous paragraph, the drop rate in our approach is $\delta=0.03$. As can be observed, the drop rate is always lower with a constant playout rate.}

\textcolor{black}{Table~\ref{table:abr} also depicts the outage probability $\epsilon$ in the three scenarios of ABR system for all the considered users, which is the third QoE metric of interest in this work. As mentioned, the outage with our system in this scenario is $\epsilon=0.01$. As can be observed from Table~\ref{table:abr}, the rebuffering events are always more often with the ABR system despite the fact that our system is more restrictive in terms of the playout rate. This is a significant advantage of our approach, as it offers both constant playout rate and less rebuffering events and loss of information.} 

%\textcolor{black}{The fourth QoE metric of interest is Eq.(\ref{eq:guohong_icdcs})~\cite{Guohong_icdcs}. As all the factors, except $I_t$, in Eq.(\ref{eq:guohong_icdcs}) are the same for our approach and the ABR, in order to simplify the comparison, we consider only $I_t$. Table~\ref{table:abr_2} depicts the results for $I_t$ in our constant playout rate approach and in the three scenarios for ABR. Note that in the QoE the factor $I_t$ has the sign ``minus'' in front of it. Hence, the lower the value of $I_t$, the higher the QoE. As can be observed from Table~\ref{table:abr_2}, our approach provides the lowest value of $I_t$ across all the users compared to the three scenarios of ABR, which means that this QoE is higher with the constant playout approach.} 

\textcolor{black}{Summarizing, we have shown that providing a constant playout rate for a maximum $\delta$ and $\epsilon$ \emph{results in a better user experience (for different QoE metrics) than allowing playout rates that are adjustable based on the state of the user's buffer}. This is one of the main messages of this paper.}  

\subsection{Impact of channel variability}
\textcolor{black}{To conclude this section, we look at the impact of channel variability on the required size of the buffer so that the loss of information (dropped packets) is very low. We pick three users with different channel characteristics: user 4 that has the lowest variation of per-block rate (its coefficient of variation is $0.03$), user 5 that has the highest variation ($c_{v,5}=0.96$), and user 1 with $c_{v,1}=0.54$. The network resources are shared equally among all the users, and the other parameters remain unchanged. We choose the (constant) playout rates to be approximately equal to the average of the data rare for each user (that is a reasonable value is expected to keep the drop rates and outages low under large buffers): $U_1=5.15$\:Mbps, $U_4=16.3$\:Mbps, and $U_5=17.1$\:Mbps. We look at what the drop rate will be as a function of the buffer size, where the latter is now expressed in seconds of live video content.}  

\textcolor{black}{Fig.~\ref{fig:buffer_seconds} shows the drop rates for the three users for different buffer sizes. User 4 (with almost constant data rates) requires a very low buffer, where even a $0.5$\:s buffer suffices to provide no loss of packets. This helps in preserving low latency (keeping the playout delay on the user's side of only $0.5$\:s). As the variability in the channel conditions increases, the size of the required buffered content increases; a buffer of $2.5$\:s is needed to yield a drop rate lower than $0.01$. Finally, for user 5, a buffer of slightly larger than $4$\:s is needed to result in a drop rate lower than $0.01$. This result is particularly important as it shows that for channel conditions with a coefficient of variation up to $1$ the latency on the user's side of less than $5$\:s is guaranteed with almost no information loss.}

\begin{table}
\centering
\caption{The outage probability and drop rate for different ABR scenarios}\label{table:abr}
%\vspace{-6pt}
\begin{tabular}{|l|l|l|l|l|l|l|l|l|} \hline
 User & 1 & 2 & 3 & 4 & 5 & 6 & 7 & 8 \\\hline
 $\delta$ (Sc.1) & 0.06 & 0.05 & 0.04 & 0.04 & 0.06 & 0.05 & 0.05 & 0.04 \\\hline 
 $\delta$ (Sc.2) & 0.06 & 0.05 & 0.04 & 0.04 & 0.11 & 0.1 & 0.09 & 0.04  \\\hline
  $\delta$ (Sc.3) & 0.11 & 0.12 & 0.08 & 0.08 & 0.15 &  0.13 & 0.12 & 0.06 \\\hline
$\epsilon$ (Sc.1) & 0.05 & 0.02 & 0.04 & 0.02 & 0.02 & 0.02 & 0.02 & 0.11\\\hline
$\epsilon$ (Sc.2) & 0.04 & 0.02 & 0.09 & 0.02 & 0.13 & 0.08 & 0.12 & 0.15 \\\hline
$\epsilon$ (Sc.3) & 0.03 & 0.02 & 0.1 & 0.02 & 0.11 & 0.07 & 0.11 & 0.17 \\\hline
 \end{tabular}
%\vspace{-10pt}
\end{table}

%\begin{table}
%\centering
%\caption{The factor $I_t$ from~\cite{Guohong_icdcs}}\label{table:abr_2}
%\vspace{-6pt}
%\begin{tabular}{|l|l|l|l|l|l|l|l|l|} \hline
% User & 1 & 2 & 3 & 4 & 5 & 6 & 7 & 8 \\\hline
% Our app. & 0.03 & 0.01 & 0.06 & 0.001 & 0.11 & 0.06 & 0.1 & 0.07 \\\hline 
% Sc.1 & 0.1 & 0.08 & 0.22 & 0.14 & 0.22 & 0.15 & 0.22 & 0.36  \\\hline
%Sc.2 & 0.04 & 0.03 & 0.12 & 0.07 & 0.14 & 0.09 & 0.14 & 0.24\\\hline
%Sc.3 & 0.08 & 0.1 & 0.35 & 0.26 & 0.28 & 0.19 & 0.29 & 0.62 \\\hline
% \end{tabular}
%\vspace{-10pt}
%\end{table}

\section{Related work}
\label{sec:related}
%In general, there are not too many works that consider consistency in 5G networks, especially not ones that focus on performance modeling and analysis of real-time video streaming. 
In general, there isn't a significant body of research dealing with resource allocation for real-time video streaming in cellular networks, especially not papers  containing an analytical framework in a 5G context. Most of the works are constrained with \emph{non-live} video streaming, such as~\cite{Hosfeld15} that deals with buffer dimensioning,~\cite{10.1145/2740070.2626296} that is a system-related paper allowing varying playout rates, or~\cite{Bulkan19} that presents an approach for predicting user's QoE. 

%\textcolor{red}{References~\cite{Nam16} and~\cite{10.1145/2018602.2018611} are studies of user viewing activities conducted on a large number of YouTube videos. In~\cite{Nam16}, the impact of different playback events, such as start-up latency, rebufferings and playout rate changes on video abandonment have been considered. One of the main findings is that frequent rebuffering in most cases is the reason for abandoning the video. On the other hand, the main outcome from~\cite{10.1145/2018602.2018611} is that video impairments lead to users pausing the video very often.}  

Network slicing as a new concept in 5G has been proposed in~\cite{Khan20} to enhance the video streaming experience in vehicular networks. The objective function combines several variables of interest. However, the playout rate is allowed to change, which deteriorates the QoE. Also, the approach in~\cite{Khan20} is not applicable to \textcolor{black}{live} video streaming.

Optimizing social welfare when streaming in real time in a mobile edge computing framework is the focus of~\cite{Hung20}. The considered metric is a special-case expression for QoE. Auctions are used to obtain the optimal solution that is not in closed-form. On the other hand, in our work we provide closed-form expressions %for two out of the three  problems 
\textcolor{black}{(whenever possible)}
corresponding to other objective functions than in~\cite{Hung20}.     

In~\cite{jigsaw}, a system, called Jigsaw, for 4K real-time video streaming is presented. While Jigsaw indeed provides a superior performance, it operates on 60 GHz frequencies, whereas our system operates in the sub-6 GHz band. Therefore, an adequate comparison between the two approaches is not feasible. Furthermore, there is no theoretical analysis in~\cite{jigsaw}.    

The related works most in spirit to ours are~\cite{Fidan_TMC} %,~\cite{Luan09}, 
and~\cite{Hsieh20}. The possibility of having a \textcolor{black}{constant} data rate that can be translated into a constant playout rate has been analyzed in~\cite{Fidan_TMC}, and the reallocation of unused resources to the same users can further improve the QoE. However, as is shown \textcolor{black}{here}, providing \textcolor{black}{constant} data rates leads to inefficient use of resources. %In~\cite{Luan09}, a theoretical framework to analyze the impact of network dynamics on the user’s perceived video quality is presented. A queueing model that relies on a diffusion approximation is used. The video quality is expressed in terms of start-up delay, fluency of video playback and packet loss. However, the paper does not capture the case of live-video streaming, which is more challenging.  
On a similar note, in~\cite{Hsieh20} the authors present a framework that relies on the Brownian approximation to optimize the QoE in terms of the playback latency and video interruptions for live streaming. %The authors propose 
A policy that jointly determines the amount of playback latency of every user and the scheduling decision of each packet transmission is proposed. It is worth noting that in our paper we follow a different approach, have different setup and objectives. The assumption in~\cite{Hsieh20} is that in a frame only one packet can be transmitted to at most one user. This is unrealistic because in 5G the frame duration is $10$\:ms,  which implies that a considerable number of packets can be forwarded. %Furthermore, in 5G the block allocation scheme~\cite{nokia5g} is used with multiple users getting videos simultaneously. 
Another simplified assumption in~\cite{Hsieh20} is that all the links exhibit similar reliability (corresponding to identical per-block rates), while we consider heterogeneous users.

\textcolor{black}{In~\cite{Xu14_buffer_thresholds}, the authors mimic the operation of DASH by introducing two levels (minimum, maximum) related to the operation of the playout buffer, and based on that determine the playout rate. On the other hand,~\cite{NOVA},~\cite{Guohong_icdcs}, provide different QoE metrics while assuming that the playout rate is variable. However, in this paper,  
we show that for several combinations of these 2-level values from~\cite{Xu14_buffer_thresholds}, our approach provides higher QoE (for any of its definitions in~\cite{NOVA},~\cite{Guohong_icdcs}) than the adaptive playout approach.}

\section{Conclusion}
\label{sec:conclusion}
In this paper, we have considered analytically the problem of resource allocation to improve %and exploiting the buffer in improving 
the QoE of cellular users with real-time video streaming, \textcolor{black}{where users have finite-size buffers.} %\textcolor{red}{It is shown that sufficiently large buffers mitigate the need for having strict constant data rates for live streaming \textcolor{red}{by allowing a slight increase in latency}.} 
We have %formulated three optimization problems to that direction, and have either 
solved three problems %them in closed-form or proposed low-complexity heuristic algorithms 
yielding the corresponding best policies. %that provide near-optimal results.  
The analysis is validated on trace data. %simulations are run on both synthetic and trace-data. 
We have compared the corresponding optimal policies against other \textcolor{black}{benchmarks}, \textcolor{black}{for different QoE metrics}. Results show that \textcolor{black}{the} performance can be improved significantly with the proper resource allocation. 
\textcolor{black}{In the future, we plan to consider resource allocation for other objectives.}

\bibliographystyle{IEEEtran}
%\bibliography{Fidan}

\end{document}